# Constraining cosmic scatter in the Galactic halo through a differential analysis of metal-poor stars

Henrique Reggiani[1, 2], Jorge Meléndez[1], Chiaki Kobayashi[3], Amanda Karakas[4], and Vinicius Placco[5]

[1] Universidade de São Paulo, Instituto de Astronomia, Geofísica e Ciências Atmosféricas, IAG, Departamento de Astronomia, Rua do Matão 1226, Cidade Universitária, 05508-900, SP, Brazil. e-mail: hreggiani@gmail.com
[2] Max-Planck Institute for Astronomy, Konigstuhl 17, 69117 Heidelberg, Germany
[3] Centre for Astrophysics Research, School of Physics, Astronomy and Mathematics, University of Hertfordshire, College Lane, Hatfield AL10 9AB, UK
[4] Monash Centre for Astrophysics, School of Physics & Astronomy, Monash University, Clayton VIC 3800, Australia
[5] Department of Physics and JINA Center for the Evolution of the Elements, University of Notre Dame, Notre Dame, IN46556, USA

**ABSTRACT**

*Context.* The chemical abundances of metal-poor halo stars are important to understanding key aspects of Galactic formation and evolution.
*Aims.* We aim to constrain Galactic chemical evolution with precise chemical abundances of metal-poor stars ($-2.8 \leq$ [Fe/H] $\leq -1.5$).
*Methods.* Using high resolution and high S/N UVES spectra of 23 stars and employing the differential analysis technique we estimated stellar parameters and obtained precise LTE chemical abundances.
*Results.* We present the abundances of Li, Na, Mg, Al, Si, Ca, Sc, Ti, V, Cr, Mn, Co, Ni, Zn, Sr, Y, Zr, and Ba. The differential technique allowed us to obtain an unprecedented low level of scatter in our analysis, with standard deviations as low as 0.05 dex, and mean errors as low as 0.05 dex for [X/Fe].
*Conclusions.* By expanding our metallicity range with precise abundances from other works, we were able to precisely constrain Galactic chemical evolution models in a wide metallicity range ($-3.6 \leq$ [Fe/H] $\leq -0.4$). The agreements and discrepancies found are key for further improvement of both models and observations. We also show that the LTE analysis of Cr II is a much more reliable source of abundance for chromium, as Cr I has important NLTE effects. These effects can be clearly seen when we compare the observed abundances of Cr I and Cr II with GCE models. While Cr I has a clear disagreement between model and observations, Cr II is very well modeled. We confirm tight increasing trends of Co and Zn toward lower metallicities, and a tight flat evolution of Ni relative to Fe. Our results strongly suggest inhomogeneous enrichment from hypernovae. Our precise stellar parameters results in a low star-to-star scatter (0.04 dex) in the Li abundances of our sample, with a mean value about 0.4 dex lower than the prediction from standard Big Bang Nucleosynthesis; we also study the relation between lithium depletion and stellar mass, but it is difficult to assess a correlation due to the limited mass range. We find two blue straggler stars, based on their very depleted Li abundances. One of them shows intriguing abundance anomalies, including a possible zinc enhancement, suggesting that zinc may have been also produced by a former AGB companion.

## 1. Introduction

The information imprinted in the chemical patterns of metal-poor ([Fe/H]$\leq -1.0$) stars hold one of the keys to understanding the formation and evolution of the Milky Way in its early stages (Eggen et al. 1962; Searle & Zinn 1978). These objects arguably offer the most powerful insights into the evolution, nucleosynthetic yields, and properties of the first supernovae, they constrain the shape of the IMF, and provide clues to the rise of the s- and r-processes in the Galaxy and the sites that produce them (Audouze & Silk 1995; Ryan et al. 1996; Shigeyama & Tsujimoto 1998; Chieffi & Limongi 2002; Umeda & Nomoto 2002; Meynet et al. 2010).

Studies of metal-poor stars are usually focused on extremely metal-poor stars ([Fe/H] $\leq -3.0$ - EMP) (e.g., Cayrel et al. 2004; Arnone et al. 2005; Cohen et al. 2008; Bonifacio et al. 2009; Hollek et al. 2011; Aoki et al. 2013; Yong et al. 2013; Jacobson et al. 2015), or in CEMP, carbon enhanced metal-poor, stars (e.g., Aoki et al. 2007; Spite et al. 2013; Keller et al. 2014; Placco et al. 2014, 2016a,b), which are the objects most likely to hold the keys to uncover details of the first generation of stars, the Pop III stars. There are also studies of the more metal-rich end of metal-poor stars (Nissen & Schuster 2010; Schuster et al. 2012; Ramírez et al. 2012; Fishlock et al. 2017), focused on stars of metallicities [Fe/H] $\geq -1.5$, which provide evidence of extra-galactic stars in the Milky Way halo.

However, there is a gap in metallicities between $-2.5 \leq$ [Fe/H] $\leq -1.5$, where there are few comprehensive studies of accurate chemical abundances and as such there are significant gaps when comparing to models of Galactic chemical evolution (e.g., Chiappini et al. (1999) for a comparison using robust statistics; Cescutti (2008) for a stochastic model; Kobayashi & Nakasato (2011) for a chemodynamical simulation). With this gap in precise abundances, model results are often compared to inhomogeneous works, obtained with different spectral resolutions and analysis methods causing large spreads in the [X/Fe] ratios, and making it very difficult for models to be properly constrained.

In this metallicity range mixing in the interstellar medium (ISM) would not have been active long enough to make all observed scatter statistical, as is the case of metal-rich stars. Thus, an extensive spread in the data would indicate the presence of real cosmic scatter and/or inhomogeneous mixing in the ISM, which could be due to the presence of different populations (as found by Nissen & Schuster (2010)). For this reason, studies of these objects can also give us important insights into the accretion of extra-galactic stars by the Milky Way.

However, to uncover such details, we must obtain precisions at the level of 0.05 dex. In order to accomplish that, we make use of the differential technique. Recently, the differential technique in twin stars (meaning stars with similar stellar parameters), made it possible to considerably improve the precision achieved in spectroscopic studies. This was possible because many error sources, such as imprecise log($gf$) values, largely cancel out, allowing a much better precision in the determination of relative stellar parameters and abundances. Studies with this technique have been used to recognize planet signatures on the chemical composition of stars (Meléndez et al. 2009; Ramírez et al. 2009; Tucci Maia et al. 2014; Biazzo et al. 2015), stellar evolution effects (Monroe et al. 2013; Tucci Maia et al. 2015; Carlos et al. 2016), chemical evolution as a function of age in the solar neighborhood (Nissen 2015; Spina et al. 2016), chemical abundance anomalies in globular clusters (Yong et al. 2013) and open clusters (Önehag et al. 2014; Liu et al. 2016b,a), distinct populations in the metal-rich halo (Nissen & Schuster 2010) and distinct chemical abundances in EMP stars by Reggiani et al. (2016). O'Malley et al. (2017) has also employed a differential analysis for an exploratory work on main sequence $-2.7 \leq$ [Fe/H] $-1.4$ stars. The abundance analysis they performed, however, is based on spectra of lower quality than in the present work, acquired using different instrumentation, and only a few elements were explored.

In this context we present a LTE differential study of the chemical abundances of 18 elements (Li, Na, Mg, Al, Si, Ca, Sc, Ti, V, Cr, Mn, Co, Ni, Zn, Sr, Y, Zr and Ba), in 23 metal-poor stars, and compare the results with a chemical evolution model, which we describe in Sect. 4.

The paper is divided as follows: in Sect. 2 we describe observations and data reduction, we detail the stellar parameters in Sect. 3, comparing our parameters to other works in Sect. 4. Chemical abundances and results are shown in Sect. 5, and lithium is studied in Sect. 6. The pair of blue straggler stars are discussed in Sect. 7 and conclusions are presented in Sect. 8.

## 2. Observations and data reduction

### 2.1. Sample selection and observations

All stars observed in this work were selected due to a proximity in their stellar parameters that allowed us to obtain precise abundances through the differential technique. Using the updated catalog of stellar parameters of Ramírez & Meléndez (2005), we selected 26 stars with previous assessments of stellar parameters within: $T_{\rm eff}$ = 6250 ± 250 K, log $g$ = 4.0 ± 0.5 dex and metallicities $-2.8 \leq$ [Fe/H] $\leq -1.5$, and brighter than V = 12, which assured that we were able to observe all stars in a reasonable time (up to two hours of exposure time) and S/N ∼ 150 - 250.

The observational data were obtained with the UVES spectrograph (Dekker et al. 2000) at the 8.2m VLT telescope, during 2015A (project 095.D-0504(A)). All the spectra were taken with the same instrumental configuration, which guarantees similar spectra quality and improves the precision in a line-by-line differential analysis. The blue side of the spectra has an effective range from 3300Å to 4500Å, and the red side of the spectra has a range of 4800Å − 6800Å. We used a 0.8" slit on both arms of the spectrograph, with a final resolution of $R \approx 50000$ per pixel in both the blue arm and red arms. The average S/N of the sample is: $S/N \approx 130$ at 4000Å and $S/N \approx 250$ at 6000Å.

Of the original 26 observed stars, we removed three from the final analysis. Two of them were too metal-rich, and one star has a very high rotation. All three were excluded from the analysis for not being compatible with the remainder of the sample.

### 2.2. Data reduction

The bias and flat field corrections, order extraction and wavelength calibration, were performed by the UVES-Echelle pipeline. Barycentric and radial velocity corrections were performed automatically via the IRAF package for python (pyraf) and the spectra normalization were performed manually for each spectra via IRAF. After the normalization process the spectra of each star were combined for the abundance analysis.

## 3. Stellar parameters

We have performed manual EW measurements, via the *splot* task in IRAF using gaussian profile fitting, for our entire sample, measuring a given line one at a time in all stars, which assures that the continuum placement of a given line is the same for all the stars, reducing the final abundance errors. We employed the semi-automatic q2[1] code (Ramírez et al. 2014), with MARCS plane-parallel 1D model atmospheres (Gustafsson et al. 2008) and the 2014 version of the LTE analysis code MOOG (Sneden 1973).

The log($gf$) values and energy levels of our linelist are from VALD (Vienna Atomic Line Database). The Fe I lines were updated using data from Den Hartog et al. (2014) and transition probabilities for the Fe II lines are from Meléndez & Barbuy (2009). The Ti II values were updated using Lawler et al. (2013). Nevertheless, we note that the choice of log($gf$) values is inconsequential in a differential analysis.

We started by performing an absolute spectroscopic measurement of the stellar parameters. Using excitation equilibrium for determining $T_{\rm eff}$, ionization equilibrium for log $g$, allowing no trend of FeI line abundances with respect to the reduced EW gave us the microturbulence ($v_T$), and using the measured EW, we obtained the initial [Fe/H] for all stars.

Analyzing the preliminary spectroscopic results we chose stars HD 338529 and CD−4802445 as our reference objects because the stellar parameters are in between the initial guess for the parameters of our other targets. We have chosen two different standard stars because the range in metallicity of our complete sample is too large. Thus, we separated the sample into two, with $-2.1 \leq$ [Fe/H] $\leq -2.7$ and $-2.1 \leq$ [Fe/H] $\leq -1.4$. We opted to use as our initial stellar parameters of HD 338529: $T_{\rm eff}$ = 6426 ± 50 K from the infrared flux method (IRFM, Meléndez et al. 2010), log $g$=4.09 ± 0.03 dex from the GAIA parallax and, using our EW, we obtained [Fe/H] = −2.29 and $v_T$=1.5 kms$^{-1}$. The initial stellar parameters of CD-48 2445 are: $T_{\rm eff}$ = 6453 ± 50 K from the IRFM (Meléndez et al. 2010), log $g$=4.23 ± 0.03 dex from the GAIA parallax and, using our EW, we obtained [Fe/H] = −1.96 and $v_T$=1.5 kms$^{-1}$.

Then, we employed a strictly line-by-line differential approach (e.g., Reggiani et al. 2016; Meléndez et al. 2012; Yong et al. 2013; Ramírez et al. 2015) to obtain the stellar parameters for the remaining targets. Using as reference the Fe I and Fe II abundances from HD 338529 and CD-48 2445, we determined $T_{\rm eff}$ through differential excitation equilibrium (e.g., Fig. 1). The $T_{\rm eff}$ have an overall good agreement with the IRFM values from Meléndez et al. (2010), when available. We obtained the log $g$ through differential ionization equilibrium, and $v_t$ by allowing no trend in the differential Fe I line abundances with reduced EW (e.g., Fig. 1), and found [Fe/H] with our line measurements. The errors for the atmospheric parameters include the degeneracy of the parameters and were determined strictly through a differential approach. The adopted stellar parameters, including errors, are provided in Table A.1.

[1] https://github.com/astroChasqui/q2

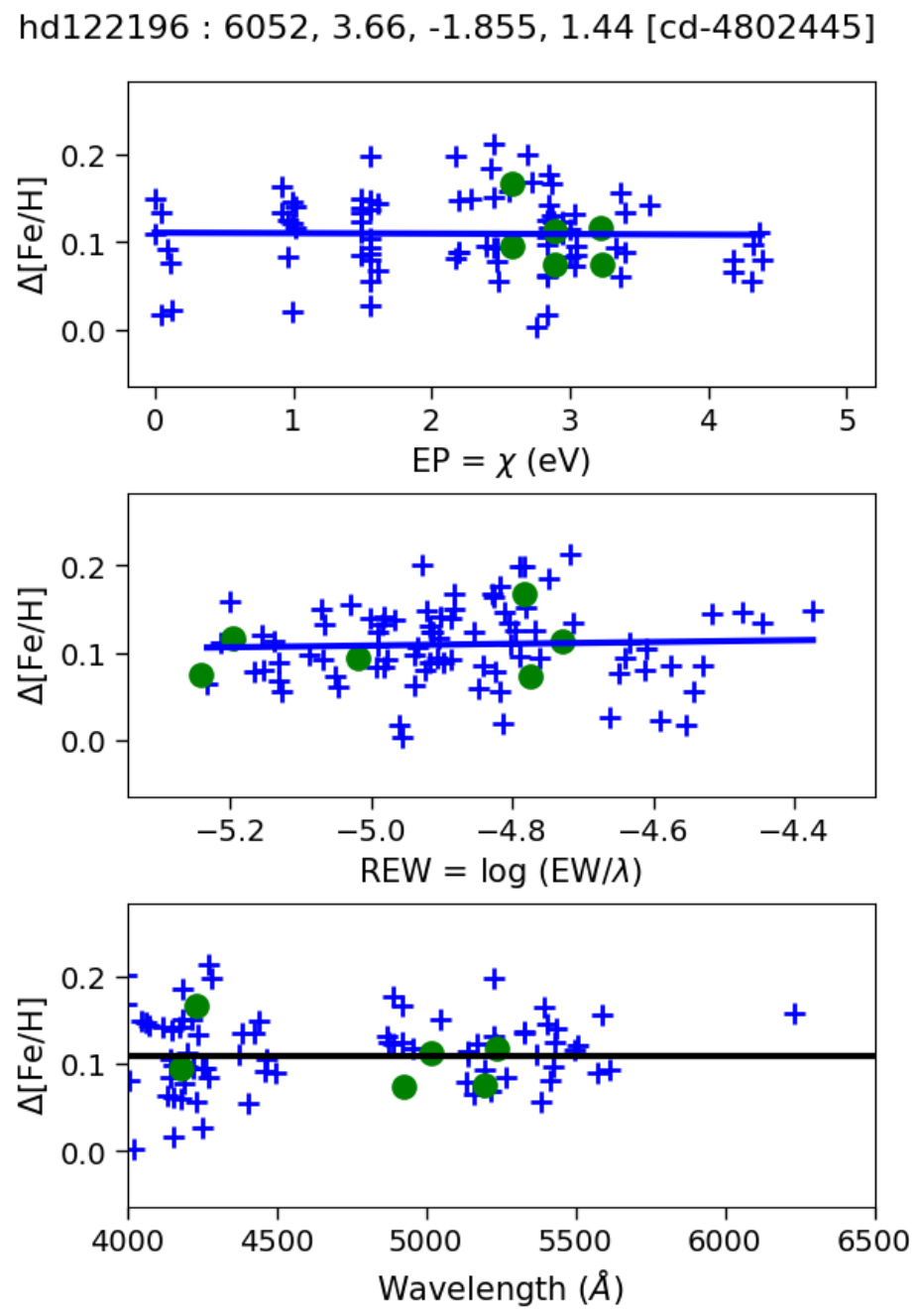

**Fig. 1.** Differential abundances in HD 122196 versus lower excitation potential (top panel), reduced equivalent widths (middle panel) and wavelength (lower panel). The blue crosses represent the differential Fe I abundances, and the green circles are the differential Fe II abundances.

## 4. Comparison with other works.

In order to check the consistency of the adopted atmospheric parameters, we have compared them to three different studies. The first is Sitnova et al. (2015) with five stars in common, the second is Bensby et al. (2014) with five stars in common, and Meléndez et al. (2010) with nine stars in common. We have calculated the difference from the atmospheric parameters ($T_{\rm eff}$ and log $g$) of each of the works cited above and our study, and calculated the median (less sensitive to the presence of outliers) of the absolute difference: $\Delta T_{\rm median}$ = 66 K, and $\Delta$log $g_{\rm median}$ = 0.18.

For the median discrepancy in temperature between our $T_{\rm eff}$ and those of the above references we find $\Delta T_{\rm median}$ = 66 K. This shows that the temperature is within a reasonable agreement between the works cited above, as the median is very similar to our calculated measurement errors and always smaller than the combination between our errors and the errors of the other works.

The median discrepancy in surface gravity is $\Delta$log $g_{\rm median}$ = 0.18 dex, which is at the upper limit of our uncertainties, but within the combined error bars of our errors and those from the literature. We note that there are differences both in the methods and data used to estimate this parameter. For example, while we used the more precise GAIA parallaxes for our standard stars, previous works used the more uncertain HIPPARCOS parallaxes. For the other stars in our sample we determined log $g$ with our differential spectroscopic approach, relative to our two standard stars, which are representative of our metal-poor sample. In this regard, we remark that other works use the Sun as a standard star, which might not be the best choice due to the large difference in stellar parameters between the Sun and such metal-deficient stars.

Sitnova et al. (2015) determined log $g$ by using as a first guess the log $g$ from Hipparcos parallaxes and adjusting them to obtain ionization balance, and Bensby et al. (2014) by applying a correction to log $g$ (from ionization balance) after a comparison between the different methods they used (which included parallaxes measurements from HIPPARCOS). Meléndez et al. (2010) determined $T_{\rm eff}$ using the IRFM, while the surface gravities are from a compilation of literature values.

The largest discrepancies are between our results and the recent work of Sitnova et al. (2015). In their work they have calculated the effective temperatures using different IRFM data sources and performed corrections of up to 50 K to obtain the ionization balance of Fe I and Fe II NLTE measurements and remove the slope in [Fe/H] vs. $E_{\rm exc}$ (excitation energies). They also employed a differential approach to calculate the parameters of their non benchmark stars, using the Sun as a standard point of comparison.

## 5. Chemical abundances

All the chemical abundances presented here are the result of a line-by-line differential analysis. The solar abundances used in this work are all from Asplund et al. (2009). We present the abundances in Figs. 2 to 6, and in Tables A.2 to A.5.

In Figs. 2 to 5 we plot the K15 model (Sneden et al. 2016; Zhao et al. 2016), which is the updated Galactic chemical evolution (GCE) of Kobayashi et al. (2011) (hereafter K11). We note that Kroupa IMF is applied in K15 and K11 models, while the Salpeter IMF is applied in Kobayashi et al. (2006). These models give almost the same results, except for C, N, Sc, Ti, V, Co, and Zn (see Sect. 5.1 for more details).

Two stars with extremely low lithium abundances, which could be blue stragglers, are indicated in the figures. Their chemical abundances have not been used to calculate the linear regression slopes of our data, the standard deviations and the mean errors presented in the figures. Although we do not use these values it is important to stress that the errors from their measurement do not increase the mean errors of the remaining sample, as they are similar to the errors of the rest of the sample.

We also added to the figures, when available, data from other precise works on unevolved stars, in order to compare observations and the galactic evolution model in a wider metallicity range ($-3.6 \leq$ [Fe/H] $\leq -0.4$). For EMP stars we add the work from Andrievsky et al. (2007, 2008); Bonifacio et al. (2009), from the First Stars large program turn-off objects, and for the more metal rich end we add data from Nissen & Schuster (2010, 2011) and Fishlock et al. (2017). We present these latter data in the figures (high-$\alpha$ and low-$\alpha$, as defined in Nissen & Schuster (2010)). We also emphasize that the above studies are based on dwarf stars of comparable atmospheric parameters as our data set.

It is important to stress that the mean values, mean errors, standard deviations, and data slopes are calculated using only our data set. The slopes calculated for the GCE model are the slopes for the $-2.7 \leq$ [Fe/H] $\leq -1.5$ region, which is the metallicity covered by our data set.

As mentioned in Sect. 3 we have separated our sample in two. We have scaled the differential abundances of Sample 2, as designated in Table A.1, based on the absolute abundances of the standard star of Sample 2, while the differential abundances of Sample 1 were scaled using the standard star of Sample 1.

### *5.1. Light even-Z metals*

The abundances of $\alpha$-elements can be seen in Fig. 2 and Table A.2.

#### 5.1.1. Magnesium

Abundances of magnesium were determined based on 8 Mg I lines, and for each star we used only lines with EW $\geq 10m$Å. As can be seen in Fig. 2 the star-to-star standard deviation of our differential [Mg/Fe] measurements is 0.06 dex, which is the same as the average error found. The small negative slope of $-0.06 \pm 0.04$ in our results, is in good agreement with the GCE prediction of a flat slope in the observed metallicity region. Previous comparisons with model data, such as the comparison in Kobayashi et al. (2006), are unable to constrain the behavior due to the large scatter in the data. However, the comparison data used in the aforementioned work is based on data from different authors, using different analyses and data sets, resulting in a large dispersion, which means it was not possible to constraint their results based on data with deviations as small as presented in this work.

To exemplify the data dispersion, we have gathered [Mg/Fe] data from the SAGA database (Suda et al. 2008, 2011; Yamada et al. 2013) with the following search parameters: $6000 \leq T_{\rm eff} \leq 6500$, $3.5 \leq \log g \leq 4.5$ and $-2.8 \leq$ [Fe/H] $\leq -1.5$, to mimic the coverage in stellar parameters of our sample. We plotted all the data returned, even the same object with several measurements, with a total of 364 data points, which can be seen in Fig. 3. Data points from the SAGA database and the stars from our sample are both plotted. We also show the linear fit to our data and the Galactic chemical evolution model prediction. As can be seen, the SAGA data agrees with the model predictions within measurement errors, in contrast to our abundance ratios which are not compatible with the GCE model. This happens because the data spread in SAGA is about 1 dex, ranging from $\approx -0.2$ to $\approx +0.8$ dex. With such a high dispersion, it is not possible to precisely constrain model results, as anything within that large range can be fitted.

The deviation between model and observations for [Mg/Fe] also extends to the works of Bonifacio et al. (2009) and Nissen & Schuster (2010), which can be seen in Fig. 2. As in our case, their sample of unevolved stars are very homogeneous. The bigger dispersion in the Bonifacio et al. (2009) results probably arise from the fact that at lower metallicities the ISM may have been more inhomogeneous. It is also interesting to see how the two populations found by Nissen & Schuster (2010) merge into one at the metal-poor end of their sample and continue without any discernible distinction from our sample toward lower metallicities. These data are important to verify the galactic chemical evolution model in a wider metallicity range. We notice that the GCE model starts to match the observational [Mg/Fe] data only at the metal-rich end.

The mean absolute value we found for Mg is [Mg/Fe]=$+0.32 \pm 0.06$ dex, which is within the expected values for the metallicity range and our low scatter is a very good improvement over what has been previously reported. Bonifacio et al. (2009) found a mean value of [Mg/Fe]= +0.21 dex with a standard deviations of 0.10 dex. This observed difference is likely due to the different samples, analyses and errors. The difference from our work and the study of giant EMP stars by Cayrel et al. (2004) is smaller. They found a mean [Mg/Fe]$\approx +0.27 \pm 0.10$ dex. But, as mentioned by Bonifacio et al. (2009), part of this discrepancy may be due to problems in line measurements of Cayrel et al. (2004), also described in Sect. 3.3 of Andrievsky et al. (2010).

Our mean [Mg/Fe]= 0.32 dex is also in agreement with the mean [Mg/Fe]= $0.29 \pm 0.07$ dex found by Zhao et al. (2016) in their NLTE analysis (in the same metallicity region). The departures from NLTE are small for Mg (Zhao et al. (2016)), hence the very good agreement with our results. They also found an offset between their NLTE abundances and the K15 GCE model of $\sim 0.25$ dex, which is the same as the mean discrepancy of our data (0.24 dex).

We note that Zhao et al. (2016) showed that [O/Fe] is consistent with the K15 GCE model at [Fe/H] $\sim -1$. This means that the model [O/Mg] is inconsistent with these observations, which is not a problem of galaxy evolution but of nucleosynthesis yields. There is an uncertainty in C($\alpha$,$\gamma$)O reaction, and the observations of stellar abundances suggest that the rate adopted in Kobayashi et al. (2006) yields (1.3 times the value given in Caughlan & Fowler (1988)) is not correct. To constrain the rate, it is necessary to use 3D and NLTE analysis for both Mg and O.

#### 5.1.2. Silicon

Our silicon abundances are based on only one measured line (3905.523Å), which is blended with CH, but is the only line that can be detected in all of our stars. Silicon has a mean error of 0.07 dex that is higher than the star-to-star standard deviation of 0.05. The calculated $0.02 \pm 0.04$ slope in our data is flatter when compared to the model prediction of $-0.06$ slope. The Galactic chemical evolution model has a plateau at about 0.6 dex and there is a large offset between the data and the model of $\approx 0.2$ dex. The mean abundance we found, [Si/Fe]= +0.16, is in agreement with the data from both Bonifacio et al. (2009) (0.09 dex) and Nissen & Schuster (2010) (0.25 dex), thus our work connects the low and high-metallicity studies.

The discrepancy between model and observations seen in this work are also observed in Zhao et al. (2016). Their mean NLTE abundances for the region $-2.6 \leq$[Fe/H] $\leq -1.4$ is +0.32 dex, which still has a considerable difference from the mean $\approx 0.59$ dex given by the model in the same region. Thus, although there are important NLTE corrections to be made in the silicon abundances, the difference between model and observations are not reconciled by the more accurate abundance estimate provided by the NLTE calculations.

#### 5.1.3. Calcium

We measured Ca abundances from CaI lines. We only considered lines with EW $\geq 10m$Å. Calcium is one of the best fits between all the data sets and GCE model predictions. As can be seen in Fig. 2 the agreement between data and predictions is impressive. It is also remarkable that the calcium slope is the same as magnesium. The GCE model agrees with almost all of our data within the error bars (0.05 dex), which are considerably lower than previous works; notice also that our error is the same as the star-to-star standard deviation of our sample. On the more metal-rich end, we see that the high-$\alpha$ population is in better agreement with the model predictions, which support the conclusion by Nissen & Schuster (2010) that the low-$\alpha$ stars might have originated in a different environment (dwarf spheroidal galaxies). At the metal-poor end we see that the data from Bonifacio et al. (2009) agrees well with the model predictions, but once again has a higher dispersion, which is not uncommon to EMP stars.

The slope we found from our data set is the same as that found for Mg, which is produced via the same mechanism as Ca, but the Ca slope has a better agreement between data and model prediction. Our mean [Ca/Fe]= 0.37 dex agrees very well with Kobayashi et al. (2006), which predicts a plateau of [Ca/Fe]≈ 0.27 – 0.39 for the metallicity range $-3 \leq$ [ Fe/H] $\leq -1$, which also matches the mean value from Arnone et al. (2005) ([Ca/Fe]= +0.37). As seen before (Bonifacio et al. 2009), we find a small difference between our work with dwarfs and the study with giants of Cayrel et al. (2004), that found a somewhat lower value in their sample. Zhao et al. (2016) reported a mean [Ca/Fe]= 0.32 ± 0.08, that also agrees with our own results and the K15 model.

#### 5.1.4. Titanium

There were approximately 50 lines of titanium measured, including Ti I and Ti II. The differential results for both species are very homogeneous (the mean values of Ti I and Ti II differ in 0.04 dex only). Our Ti data has a slope of −0.13 ± 0.04 and the galactic evolution model predicts a slope of −0.04. There is also an offset of ≈ 0.4 dex between model and our data. Even larger discrepancies had already been seen in Kobayashi et al. (2006), and the effects of jet-like explosions to enhance Ti, first proposed by Maeda & Nomoto (2003), has been applied in the K15 GCE model plotted here. But these effects are not enough to remove the discrepancy with the observed data.

The mean value (averaging Ti I and TiII) is [Ti/Fe]=0.41 dex. The overall behavior of Ti, also considering the data sets from Bonifacio et al. (2009); Nissen & Schuster (2010), is a decrease in [Ti/Fe] with an increase in metallicity. At the more metal-rich end the model seems to agree with the low-$\alpha$ population, which is in contrast with what is seen in [Ca/Fe]. If we are to interpret the low-$\alpha$ as a population from another environment, we should expect that [Ti/Fe] to be in agreement with the high-$\alpha$ population, born in the Milky Way. Thus, the model and observations of Ti also do not match at the more metal-rich end, although the discrepancy is smaller.

The discrepancy between observations and the K15 model also extends to the NLTE analysis of Zhao et al. (2016), although they found a somewhat smaller abundance [Ti/Fe]= 0.30 ± 0.05, decreasing the discrepancy with the model. We emphasize that, within the errors, the results of Zhao et al. (2016) are compatible with ours.

The results presented in Fig. 2 do not indicate the presence of any extra-galactic objects or different populations other than regular Milky Way metal-poor stars in our sample of very metal-poor objects. However, it is important to stress that the star-to-star scatter in our abundances is similar to the abundance errors, thus, to fully discard the presence of separate populations, it is necessary to obtain better data, with higher S/N and spectral resolution, which will improve the errors to a level below the current observed star-to-star scatter, hence bringing tighter constraints on the true cosmic scatter in metal-poor stars. The comparison between GCE model and observations show that there still is a discrepancy of ∼ 0.3 dex in Mg, Si, and Ti predictions, but Ca nucleosynthesis seems to be very well defined in the K15 model, matching the observations from EMP to almost solar metallicity stars.

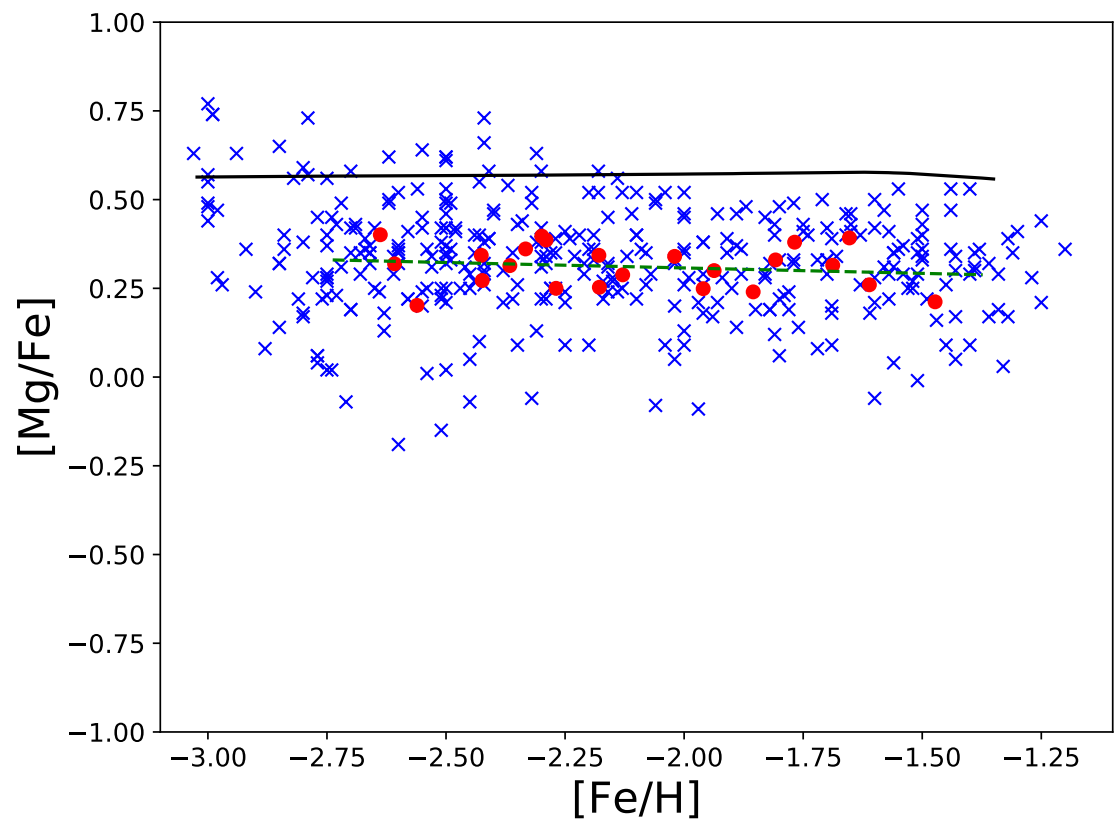

**Fig. 3.** [Mg/Fe] abundances from the SAGA database. Notice the big data dispersion that arises from different analysis methods and samples (blue crosses). The stars selected from the SAGA database have $T_{eff}$ = 6250 ± 250 and log $g$ = 4.0 ± 0.5, as in our sample. For comparison our more precise results are shown by red circles. The black line is the GCE model prediction.

### 5.2. Light odd-Z metals

All the odd-Z light elements results can be seen in Fig. 4 and Table A.3.

#### 5.2.1. Sodium

The LTE differential results have the largest star-to-star standard deviation among the light elements (0.17 dex) and a mean error of 0.09 dex with a slope of +0.35 ± 0.09, much higher than the 0.17 predicted by the GCE model. Such high star-to-star standard deviation and steep slope had already been reported by other authors (e.g., Cayrel et al. 2004). The absolute mean value we found [Na/Fe]= +0.22 dex, is much higher than the −0.07 dex predicted by the GCE model.

It is well known, however, that there are considerable NLTE departures for sodium that can change the estimated abundances up to 0.5 dex (Baumueller et al. 1998), especially on the resonant 588.9 and 589.5 nm lines that were used to determine the LTE differential abundances. We corrected our abundances using the Lind et al. (2011) data, available trough the Inspect[2] project. These corrections were done to each of the resonant lines, which suffer large NLTE effects and saturate on the more metal-rich stars. After taking the useful suggestions from the referee we also added measurements of the 568.2 and 568.8 nm lines (which are less sensitive to NLTE departures) when the lines were clearly measurable - stars with [Fe/H]≥ −2.1 (except for a couple of stars where the lines were not detected due to bad S/N in that region of the spectra). The final [Na/Fe] abundances are listed in Table A.3 and shown in Fig. 4.

With the NLTE corrections, the behavior of our sample is in good agreement with the data from other works, making a solid "bridge" between the EMP and MP stars, and the slope ($-0.02 \pm 0.08$) is in better agreement with the model. The star-to-star standard deviation of 0.12 dex is also similar to the scatter of 0.13 dex found by Andrievsky et al. (2007). Since Na production highly depends on the initial metallicity of progenitor stars, it is very unlikely to have a negative slope in the [Na/Fe] evolution. It is also important to see how the observations begin to deviate from the model for metallicities [Fe/H]$\geq -2.2$, probably due to problems in the NLTE treatment of the data. Better measurements of the weaker 568 nm lines in all the stars could possibly alleviate the issue.

[2] http://inspect-stars.com/

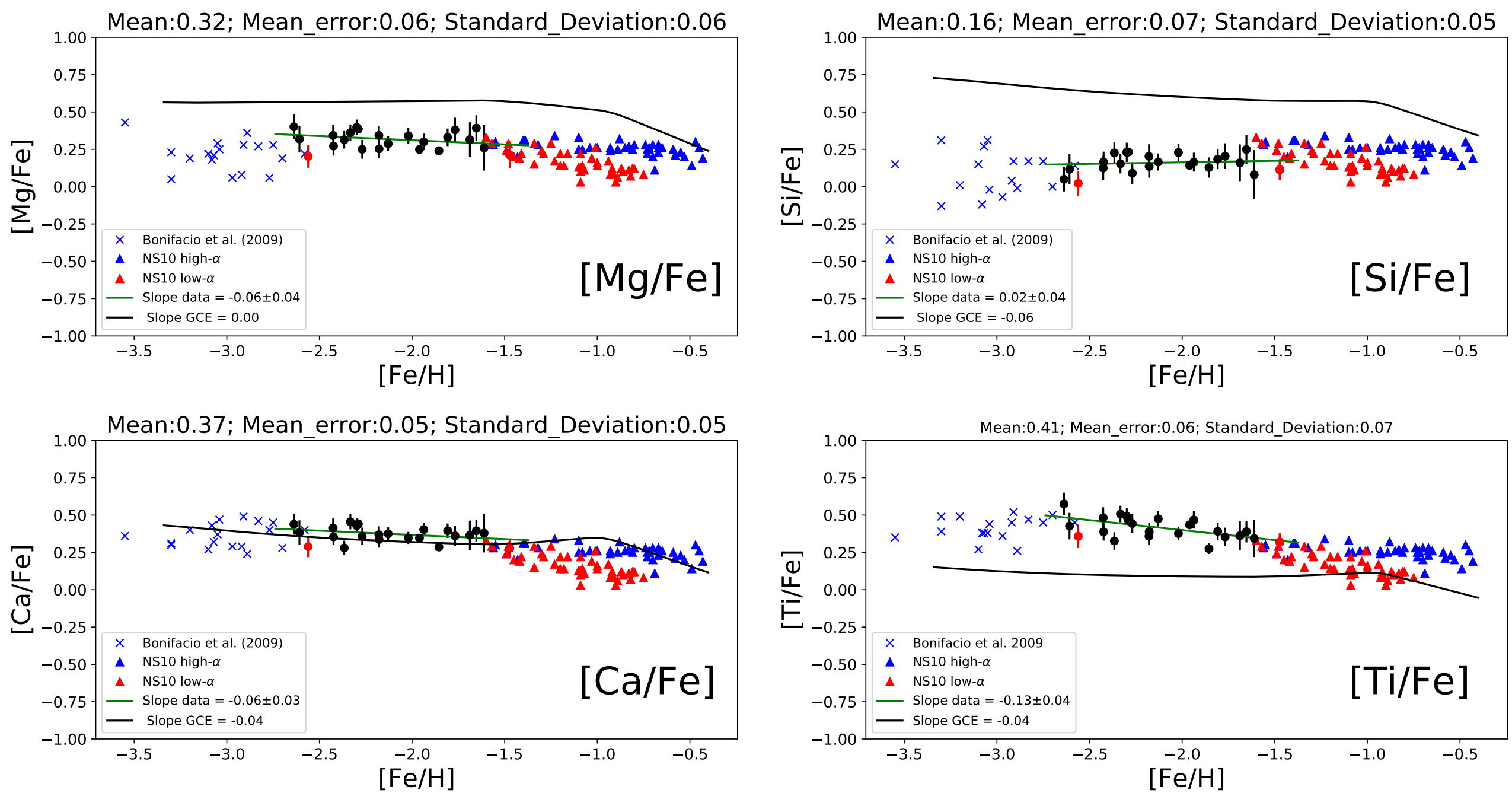

**Fig. 2.** [X/Fe] abundances for $\alpha$-elements (Mg,Si,Ca and Ti). The green line represent the best linear fit to the data and the slope is shown in the plots. The black line is the GCE prediction. The mean differential, our errors and the standard star-to-star scatter (standard deviation) for each element (for our measurements) are shown in top of each panel. The blue crosses are data of EMP stars from Bonifacio et al. (2009), the blue and red triangles are metal-poor stars from Nissen & Schuster (2010), the black filled circles are the data measured in this work and the red filled circles are the blue straggler stars from this sample.

The considerable NLTE corrections had a major effect on the mean [Na/Fe]. The LTE mean is [Na/Fe]= +0.22 dex, while the NLTE result is [Na$_{NLTE}$/Fe]= $-0.22$ dex. In contrast to the halo data shown in Kobayashi et al. (2006), the scatter in our results are small. It is also worth mentioning that with the applied corrections the scatter in our data is smaller, which might validate the adopted NLTE corrections.

Sodium abundances from Zhao et al. (2016) are well reproduced by the K15 model, but their observations have a larger scatter than what we find here. We also see the Na overproduction for higher metallicities in the K15 model, when compared to their measurements, also seen in the [Na/Fe] shown here, from Nissen & Schuster (2010). We note the Na overproduction of AGB stars has been solved in the Karakas (2010) yields, and this problem is likely to be caused by the metallicity dependence of core-collapse supernovae.

### 5.2.2. Aluminum

As for sodium, NLTE effects play an important role in aluminum abundances in metal-poor stars. It can be seen from the lower panel of Fig. 2 of Andrievsky et al. (2008) that in the temperature, surface gravity, and metallicity range of our stars, the NLTE corrections are almost constant (they range from 0.6 to 0.7 dex). This is a similar result to what was seen in the correction grid provided by Baumueller & Gehren (1997). Thus, we applied the same correction of +0.65 dex to all our stars, which left the differential abundances, and errors, unaltered. However, it is important to use these corrections with caution, because as pointed by Andrievsky et al. (2008), the shapes of the LTE and NLTE profiles are different, therefore spectral synthesis is more appropriate than NLTE corrections. As we did not have access to the NLTE spectral synthesis of Al, we chose to use NLTE corrections to assess their effect on the GCE.

The Al data scatter and the mean error are on the same level (0.09 and 0.08 dex), which indicates a very small, if any, astrophysical scatter for Al in our sample. The scatter we found is similar to the low standard deviation of the turnoff objects published in Andrievsky et al. (2008) (0.09 dex), both our work and that of Andrievsky et al. (2008) have a lower scatter than the LTE work by Cayrel et al. (2004) ($\sigma_{Al}$ = 0.21).

When compared to the GCE model, there is a small disagreement. In Fig. 4 we see that the data from Andrievsky et al. (2008) agrees with the model prediction until [Fe/H]$\sim -3$. For higher metallicities there is a disagreement of up to 0.25 dex and the overall behavior is a flat slope. Notice that the more metal-rich stars from Andrievsky et al. (2008) are in good agreement with our data. Albeit there is discrepancy between the model predic-tions and our data, which fall lower than predicted, a more proper NLTE approach should be followed to confirm this behavior. Although there is a discrepancy, our results have a much smaller scatter than previous works and, along with quality data on different metallicities ranges, will be important to improve GCE models.

Zhao et al. (2016) found a similar result in his NLTE analysis. The Al trend in the K15 model is similar to the Na trend, and there is an offset of ~ 0.25 dex between data and model. Our offset is slightly higher (0.29 dex), but the overall behavior of the data is similar and supports the previous work. The mismatch of [Na/Al] at low metallicity might also be related to the mismatch of [O/Mg] ratios.

#### 5.2.3. Scandium

Our Sc abundances were measured using Sc II lines. We corrected the lines for hyperfine structure using HFS data from Kurucz[3] linelists. The scatter in our sample is 0.07 dex, which is the same as reported by Cayrel et al. (2004) and also very similar to the ~ 0.1 dex standard deviation on Bonifacio et al. (2009). One could expect a higher scatter for Sc abundance, compared to other elements, due to the fact that the nucleosynthesis of this element is heavily dependent on the mass of the progenitor (Chieffi & Limongi 2002). We, however, find the scatter of Sc to be at the same level as in the other light elements. The overall behavior of our data points is consistent with previous results from Cayrel et al. (2004) and follow closely what was found by Bonifacio et al. (2009) and seems to be quite in good agreement with the trend they found.

It is intriguing to observe that both Bonifacio et al. (2009) and our work have [Sc/Fe] ratios significantly higher than the metal-rich halo stars studied by Fishlock et al. (2017), which are from the high and low-$\alpha$ populations from Nissen & Schuster (2010). Their analysis was based on spectral synthesis and $\chi^2$ minimization of one Sc II line, while ours was performed via analysis of the curve of growth, both using MOOG (Sneden 1973). Although at first sight we could think of systematic errors, our curve of growth LTE analysis of Sc in a star from the Nissen & Schuster (2010), shows that indeed the metal-rich halo stars have [Sc/Fe] ratios slightly lower than solar (Reggiani & Melendez 2017, in prep).

There is a disagreement between scandium measurements (ours and Bonifacio et al. (2009)) and the galactic chemical evolution model, but such a difference was already reported in Kobayashi et al. (2006). In that work there is a difference of almost 1 dex, while the difference with our data is about 0.7 dex. This is likely due to the effect of jet-like explosions applied to the K15 model (Sneden et al. 2016). Sc yields could be more enhanced by the $\nu$ process (Kobayashi et al. 2011), which is not included in the K15 model. Interestingly the GCE model agrees better with the data from Fishlock et al. (2017) for the more metal-rich stars.

The [Sc/Fe] data by Zhao et al. (2016) follows a similar behavior to our data, Bonifacio et al. (2009) and Fishlock et al. (2017), showing a disagreement with the K15 model at low metallicities and an agreement for the metal-rich objects. However, the Zhao et al. (2016) scandium abundances are lower ([Sc/Fe]~ 0.2 dex versus our [Sc/Fe]~ 0.3 dex), and do not have a big offset between the more metal-poor and more metal-rich stars, thus having a somewhat smaller difference when compared to the K15 model. That indicates that a NLTE treatment is more accurate for Sc measurements for more metal-poor stars.

[3] http://kurucz.harvard.edu/linelists.html

### 5.3. Iron-peak

The results for the iron-peak elements are presented in Fig. 5, and in Table A.4.

#### 5.3.1. Vanadium

Vanadium abundances were calculated from V II lines with hyperfine structure data from Wood et al. (2014). There are only a few abundances of V for halo stars in Kobayashi et al. (2006) and the data points are scattered, which shows the difficulty in comparing model results with actual data. Our results indicate a slope of −0.25 ± 0.06, that is steeper than the GCE predictions. The data results are also somewhat higher than the GCE model, as previously seen in Kobayashi et al. (2006) and in the K15 model with the effects of jet-like explosions. As for Sc, V yields could be enhanced by the $\nu$ process (Kobayashi et al. 2011). The mean error and scatter are on the same level. It is puzzling the extremely lower V abundance of one of the blue straggler stars; this will be discussed further in Sect. 7.

#### 5.3.2. Chromium

In previous works, such as Cayrel et al. (2004), Cr is found to have a positive slope, meaning a decreasing abundance with decreasing metallicity. Bonifacio et al. (2009) also found a similar behavior for their turnoff stars (see their Fig. 8). Our Cr I data is consistent with that behavior but presents a steeper slope (0.40 ± 0.06 against a 0.12 reported by Cayrel et al. (2004)). All the chromium results from Cr I (Bonifacio et al. (2009), this work, and Nissen & Schuster (2010)) are inconsistent with GCE model predictions and our Cr I abundances vary from lower to higher than the model, for the more metal-poor and more metal-rich end of our sample, respectively.

The behavior of Cr II is, however, very different. In our sample the slope of Cr II is 0.01 ± 0.03, which is consistent with the −0.01 GCE model results, as can be seen in Fig. 5. This difference between CrI and CrII measurements had already been reported in Kobayashi et al. (2006); Lai et al. (2008); Bonifacio et al. (2009). As in Kobayashi et al. (2006), we consider the LTE analysis of Cr II to be better than Cr I in LTE, to trace the chemical evolution of this element. The star-to-star standard deviation of Cr II is smaller than the errors (0.04 dex and 0.07 dex respectively) and this scatter is among the lowest of our sample. In Fig. 5 we show CrI and CrII abundances in different panels, for our data and those of Bonifacio et al. (2009) and Nissen & Schuster (2010), together with the GCE model, which predicts roughly a flat plateau in [Cr/Fe]. It is clear than the agreement is better for Cr II. The lesser agreement for Cr I is probably due to NLTE effects.

#### 5.3.3. Manganese

Manganese abundances were calculated using hyperfine components from Kurucz[1]. In Fig. 5 we can see that the mean error and the scatter for Mn are 0.07 and 0.09 dex, suggesting the scatter might have an astrophysical origin. Our results are in agreement with the star-to-star scatter in the metal-poor giants of Cayrel et al. (2004). The steep slope we found for Mn (0.23 ± 0.03) is much steeper than GCE model prediction, but it seems to be in good agreement with data from Bonifacio et al. (2009); Nissen & Schuster (2011). Thus, our [Mn/Fe] measurements connect well with lower and higher metallicity data.

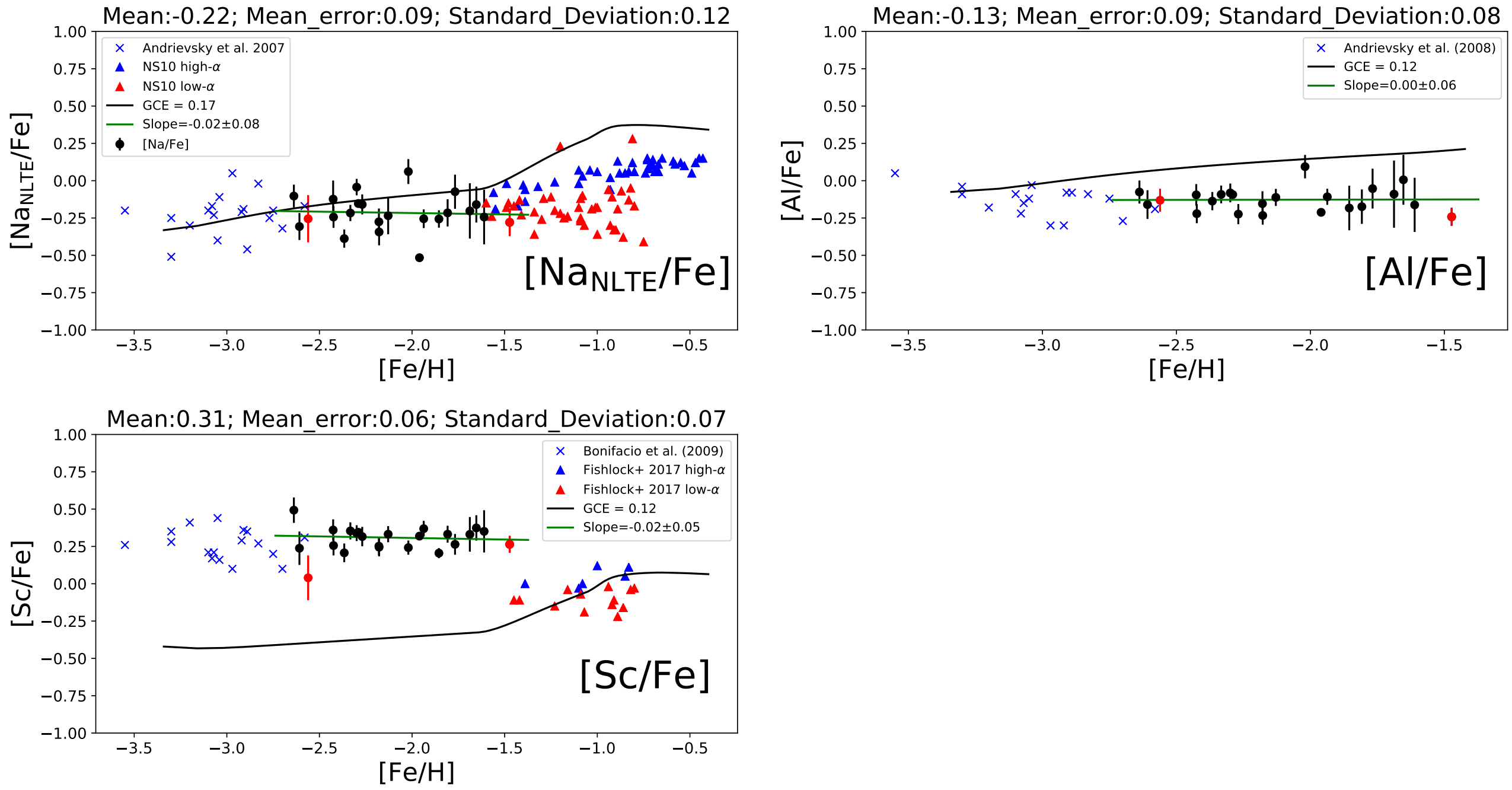

**Fig. 4.** Same as Fig. 2 for light odd Z elements (Na, Al and Sc). The data for EMP stars (blue crosses) for Na, Al, and Sc come from Andrievsky et al. (2007, 2008); Bonifacio et al. (2009) respectively, the data for the metal rich end for Na, and Sc (triangles) are from Nissen & Schuster (2010); Fishlock et al. (2017) respectively, the black filled circles are the data measured in this work and the red filled circles the two blue straggler stars from this sample.

Manganese is an odd-Z element and Mn yields depends on the progenitor metallicity, but such a steep increase in abundance with increasing metallicity was not seen in the GCE model. When comparing our data with the scattered data plot seen in Kobayashi et al. (2006) (see their Fig. 22) one can notice that their data also has higher values of Mn when compared to the model, but the scatter is big and makes it difficult to assess whether the model agrees or not with those earlier literature values.

### 5.3.4. Cobalt

We have also used Kurucz's HFS data for Co abundances. We see in Fig. 5 that the cobalt abundances decrease with increasing metallicity throughout the entire metallicity range being analyzed. Our data have a steep slope of $-0.28 \pm 0.03$, which is steeper than the $-0.02$ value predicted by GCE model. The disagreement between data and GCE model is also seen in Kobayashi et al. (2006). Our errors are at the same level as the scatter, but our data is more precise than previous works.

### 5.3.5. Nickel

Although nickel is synthesized in the same process as Co, in the complete Si burning region, Cayrel et al. (2004); Lai et al. (2008); Bonifacio et al. (2009) had already reported that while [Co/Fe] decreases with increasing [Fe/H], [Ni/Fe] remains flat. We also found a flat slope for nickel ($0.00 \pm 0.02$), which is consistent with the 0.02 predicted by the GCE model, and data from different metallicity ranges. There is an impressive flat plateau for the [Ni/Fe] ratios between $-3.6 \leq$[Fe/H]$\leq -0.4$, indicating a very homogeneous nickel production throughout cosmic history. It is also important to stress that the [Ni/Fe] star-to-star scatter (0.04 dex) is significantly smaller than the error (0.08 dex).

### 5.3.6. Zinc

Zinc is also mainly produced in the complete Si burning region, but can also be produced in neutron capture processes in more metal-rich stars (Kobayashi et al. 2006), and there is a negligible portion of Zn being produced in electron-capture supernovae (Kobayashi, Karakas, et al., in prep.). Depending on the neutrino physics, Co can be enhanced instead of Zn by electron-capture supernovae (Pllumbi et al. 2015). Both Cayrel et al. (2004) and Bonifacio et al. (2009) found a slope similar to previous results, indicating a formation processes consistent with complete silicon burning. We see the same behavior as in Co (complete Si burning), with a steep slope ($-0.16 \pm 0.05$) against a flat model prediction, going from the Bonifacio et al. (2009) data, all the way to a metallicity of [Fe/H]$= -0.4$. When we consider our data set alone there is one data point (the more metal-poor Blue Stragler) that has a higher abundance, and being a blue straggler this effect could be interpreted as the result of a possible different nucleosynthetic origin. However, when considering also the data from Bonifacio et al. (2009), it seems that the higher zinc abundance of this object is just an effect of cosmic scatter. More data on blue stragglers (BSS) zinc abundances are necessary in order to say if the higher abundance of this star has anything to do with the BSS phenomena.

It is important to note that this GCE model is a so-called one-zone model where instantaneous mixing is assumed. This assumption is valid probably for [Fe/H]> −2, but not for [Fe/H]< −2.5 where chemical enrichment should take place inhomogeneously and EMP stars are enriched only by one or two supernovae (Audouze & Silk 1995). The increasing trends of Co and Zn (and the flat trend of Ni relative to Fe) may be explained more realistically, via chemodynamical simulations (e.g., Kobayashi & Nakasato 2011). From a nucleosynthetic point of view, both [Co/Fe] and [Zn/Fe] increase with higher explosion energy (i.e., hypernovae, Kobayashi et al. 2006) and it is possible to predict some variation in Co/Zn. With higher energy, the Fe production mass is larger, but because of the larger amount of H mixed into the ejecta, the [Fe/H] of the EMP stars can be smaller (Nomoto et al. 2013). Our tight trend of Co and (less tight) trend of Zn is suggestive of inhomogeneous enrichment from hypernovae. Ni/Fe does not depend on the explosion energy nor on mass very much, and the flat trend with the small scatter gives strong constraints on the mixing-fallback mechanism of core-collapse supernovae (both for supernovae and hypernovae).

### 5.4. Neutron-capture elements

The abundances of the heavy elements Sr, Y, Zr and Ba can be seen in Fig. 6, and are shown in Table A.5. The GCE model we have been using to compare our data does not go further than Zn. Heavier elements are predominantly produced by neutron capture events (Meyer 1994). The two main neutron capture processes are the rapid neutron capture process (r-process) and the slow neutron capture process (s-process; Busso et al. 1999; Karakas & Lattanzio 2014), where most of the s-process production occurs in low-mass AGB stars (Busso et al. 2001). The s-process can also occur in fast rotating massive stars (Pignatari et al. 2010; Frischknecht et al. 2016), which may have an important contribution at low metallicity. The yields of s-process elements depend on masses and initial compositions of these stars, and the result of these processes can be observed as cosmic scatter for more metal-poor stars, formed before the interstellar medium properly mixed the material, and a more statistical scatter for the more metal-rich stars where mixing in the ISM had more time to work. While there are uncertainties surrounding the details of the s-process, the site is reasonably well understood, in contrast to the r-process. The origin of the r-process is unknown and it could occur in different sites, such as SNe II or neutron stars mergers (Cowan & Sneden 2004; Thielemann et al. 2011; Ji et al. 2016). There is also a third possible mechanism to produce heavy elements, an intermediate neutron capture process, which takes places in neutron flux densities between the s and r processes, called the i-process (Cowan & Rose 1977; Hampel et al. 2016). There are evidences of i-process in the metal-poor stars nucleosynthetic history (Herwig et al. 2014; Roederer et al. 2016).

These uncertainties in the production sites of i-process and r-process elements, along with a limited number of published yields of s-process from metal-poor stellar models limit the capability of modeling such elements. All these difficulties increase the importance of precise chemical abundances of as many stars as possible with broad wavelength coverage.

#### 5.4.1. Strontium

Strontium abundances were calculated from two Sr II lines, which are not significantly affected by NLTE effects according to Hansen et al. (2013). The authors show that accurate Sr II abundances can be obtained if reliable effective temperatures and surface gravities are available, such as in our case. The steep positive slope is mainly due to two more metal-poor stars that appear to have an extremely lower Sr abundance. This lower abundance, almost 1 dex for the most Sr deficient star, could suggest that this star was formed in an environment where AGB stars had not been activated yet, which would greatly decrease the s-process element production and become apparent in its abundance pattern. However, it is important to stress that the scatter in Sr is the second biggest in our measurements, lower only to the scatter in barium. The scatter becomes more clear when considering also the data from Bonifacio et al. (2009), which allows us to see that the two low Sr stars in our sample are probably just other examples of the very big large spread in [Sr/Fe]. This scatter has been previously reported (McWilliam 1998; Cayrel et al. 2004; François et al. 2007; Lai et al. 2008; Bonifacio et al. 2009) and was confirmed in the NLTE analysis of Andrievsky et al. (2011). As pointed out by Andrievsky et al. (2011), the scatter of strontium decreases at higher metallicities, which agrees with our observations. Overall, the data suggests that the scatter in [Sr/Fe] decrease for [Fe/H] > −2.4. Unfortunately, even with high precision data available, the current errors on stellar yields do not allow us to draw conclusions about the chemical evolution of strontium in the early Galaxy (Hansen et al. 2013), and the nucleosynthetic sites in which it might be produced.

#### 5.4.2. Yttrium

According to Hannaford et al. (1982): "the effects due to isotopic splitting and hyperfine structure in yttrium are insignificant, because there is only one stable isotope, and the hyperfine splitting is very small, typically less than 1 mA". Thus, yttrium abundances were calculated from five YII lines, without hyperfine or isotopic corrections. As with Sr, there is significant scatter. The slope is almost flat, but with a big uncertainty, and we see one star with much lower [Y/Fe], which is the same object that deviates almost 1 dex in Sr, showing that this star indeed does have lower s-process abundances.

The scatter in Y abundances also extends to the higher metallicity sample of Nissen & Schuster (2011), but there is a very well defined separation between their low and high-$\alpha$ populations. It is difficult to assess if our data follow a similar behavior because the high scatter we observe is only present in the neutron-capture elements, not the $\alpha$-elements, as seen in Nissen & Schuster (2010, 2011). Also, the scatter seems largest for [Fe/H] < −2.4.

#### 5.4.3. Zirconium

Zirconium abundances were obtained from three Zr II lines and it has the smallest deviation among the heavy elements in this work. The calculated slope is negative, but the mean error is closest to the star-to-star standard deviation than any other heavy element.

Among the neutron-capture elements, Zr is the element that deviates the most from the more metal-rich sample of Fishlock et al. (2017). Their data show a much lower mean abundance of this element, and differently from the other neutron-capture elements, it does not seem to be a connection between their higher metallicity sample and our lower metallicity range. It is unclear if this is a result of the nucleosynthetic history of the element or due systematic differences in the analyses.

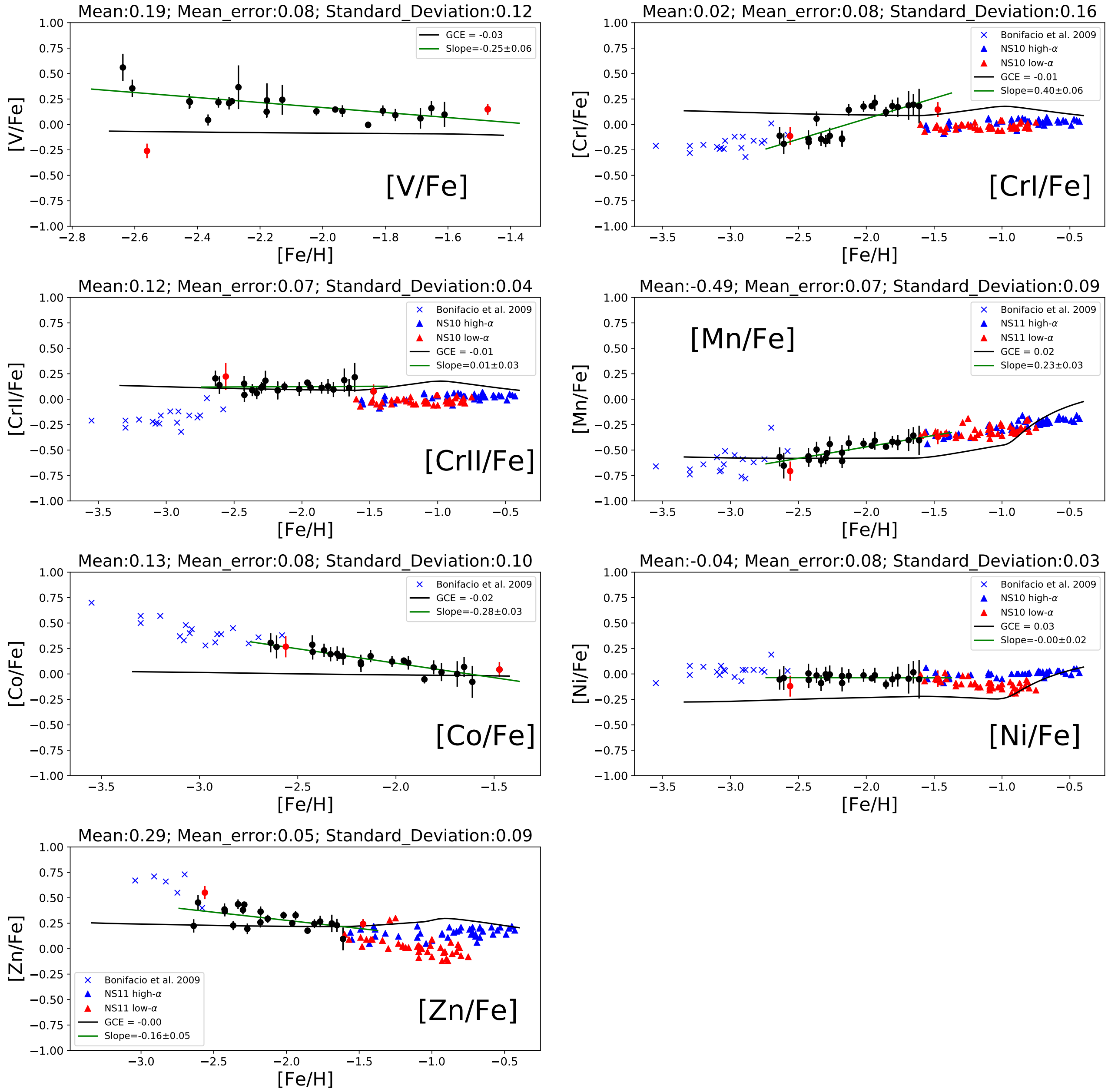

**Fig. 5.** Same as Fig. 2 for iron peak elements (V, Cr, Mn, Co, Ni and Zn). The blue crosses are data from Bonifacio et al. (2009), the red and blue triangles are the Nissen & Schuster (2010, 2011) measurements, the black filled circles are the data measured in this work and the red filled circles the two blue stragglers stars from this sample.

#### 5.4.4. Barium

The last heavy element analyzed in our sample is barium. We have applied isotopic splitting corrections from McWilliam (1998). In the solar system Ba is mainly produced via the s-process (85%, McWilliam 1998), while the remainder is produced via the r-process. However, this production scenario can be different for metal-poor stars, where the r-process might have more significant contribution. Our results indicate a very steep slope for Ba, not consistent with the other s-process elements. This is mainly due to differences in the most metal-poor end of our sample, which have consistently lower abundances, perhaps bringing insights on s-process nucleosynthesis. Consider-ing the large scatter observed, our [Ba/Fe] ratios are consistent with the metal-poor sample from Bonifacio et al. (2009) and partly consistent with the metal-rich sample by Nissen & Schuster (2011), albeit most of their sample seem to group around [Ba/Fe] $\sim -0.2$.

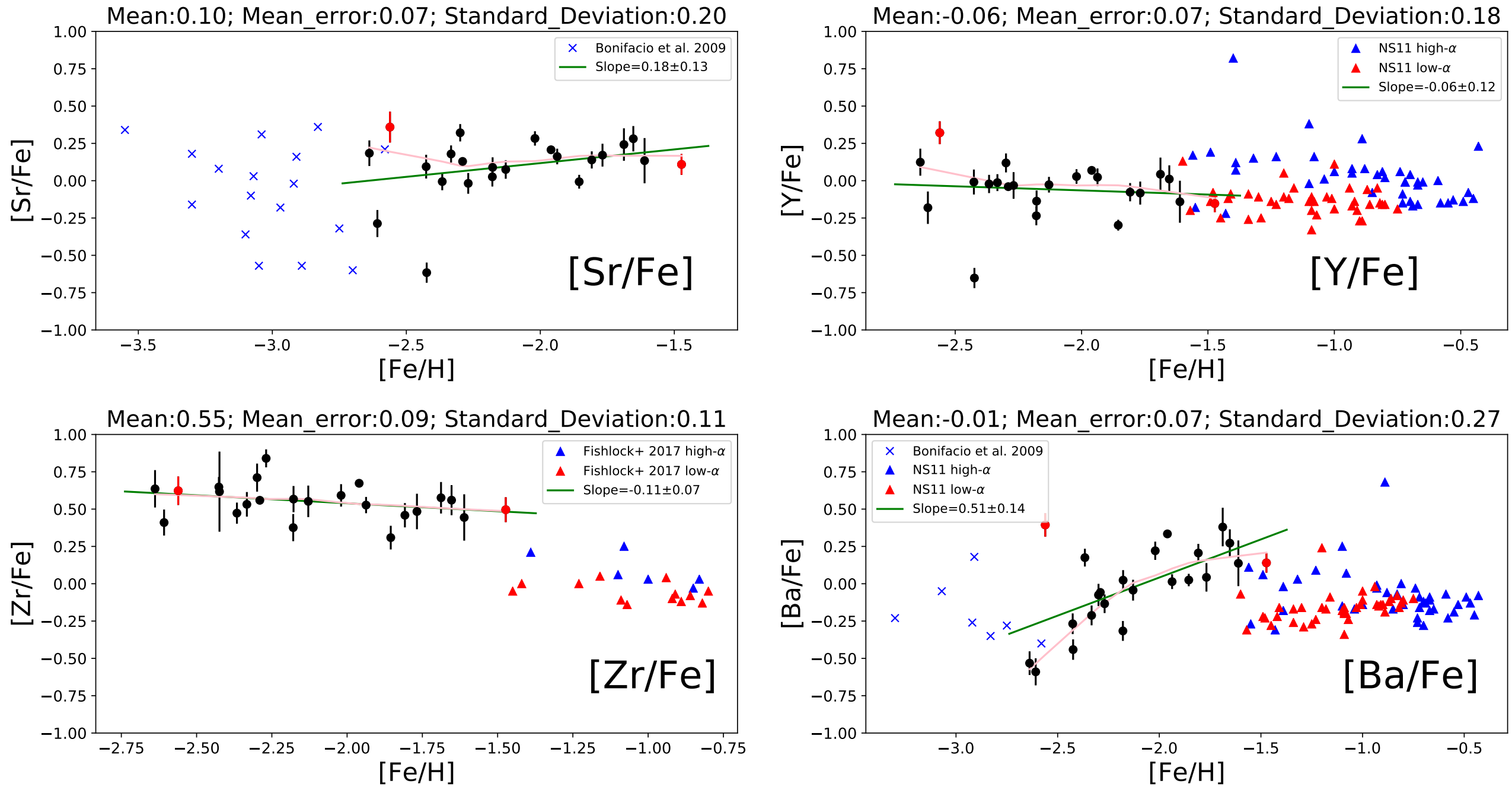

**Fig. 6.** Same as Fig. 2 for neutron-capture elements (Sr, Y, Zr and Ba). For Sr and Ba we show the Bonifacio et al. (2009) data for EMP stars as blue crosses, the blue and red triangles on Y, Ba and Zr are from Nissen & Schuster (2011); Fishlock et al. (2017) respectively, the black filled circles are the data measured in this work and the red filled circles the two blue stragglers stars from this sample.

This rather large scatter in [Ba/Fe] could be partly due to NLTE effects. As pointed out in Andrievsky et al. (2009), the NLTE corrections in this metallicity regime rapidly increase with increasing temperature. Thus, even in a homogeneous sample like ours there might be important NLTE corrections in barium abundances. Andrievsky et al. (2009) showed that, even with the NLTE calculations, there is considerable scatter in barium abundances, which support a complex evolution throughout cosmic time, with the possible additional contribution of the r-process (e.g., François et al. 2007). It is also important to stress that the low Sr and low Y star, also has a lower Ba abundance, compared to the other object with the same [Fe/H], although the difference is not as considerable.

The [X/Fe] abundances of heavy elements in our sample show a very big dispersion. The star-to-star scatter are greater than all the mean errors and also much higher than in the other elements studied. This higher scatter, and diverse nucleosynthesis origins, indicate that the results point to cosmic scatter. This suggests that a linear regression might not be the best function to describe the chemical evolution of these elements. Thus, we added a non-parametric regression to our data set, which can be seen as the pink lines in all panels of Fig. 6. We used a LOWESS function, which fits simple models to localized subsets of data, models that are used to build the function that best describes the variation in the data, point by point. The LOWESS regression works better with large data sets, but we have applied it to our sample of heavy element abundances in order to see the difference between a local regression and a linear regression.

The LOWESS function, Fig. 6, indicates that where the interstellar medium had more time to mix the materials (the more metal-rich end) the linear regression and the non-parametric regressions are close to each other and, as the metallicity decreases the two regressions deviate. Our smaller error bars and precise abundances will be important to constrain the nucleosynthesis of these elements and will help to constrain the rise of the s-process in the Galaxy.

## 6. Lithium

Spite & Spite (1982) found that warm metal-poor stars have a constant lithium abundance, and interpreted their finding as relic lithium from primordial nucleosynthesis. However, Big Bang nucleosynthesis (BBN) predictions, along with baryon density observations from Planck, predicts A(Li)=2.67 (Coc et al. 2014b; Cyburt et al. 2016), which is $\approx 0.4$ dex higher than what is observed in metal-poor stars (e.g., Spite & Spite 1982; Asplund et al. 2006; Bonifacio et al. 2007; Meléndez et al. 2010; Spite et al. 2015). This discrepancy has been the the focus of many different studies over the years. Possible explanations include new physics (e.g., Coc et al. 2009; Iocco et al. 2009; Kohri & Santoso 2009; Civitarese & Mosquera 2013; Coc et al. 2014a; Salvati et al. 2016; Hou et al. 2017) and stellar evolution effects (Richard et al. 2005; Fu et al. 2015).

Our lithium abundances are computed in NLTE, following Lind et al. (2009), and are presented in Fig. 7. The stellar masses from Fig. 8 were estimated using the q2 code, which uses Y2 isochrones (Yi et al. 2001; Kim et al. 2002) to the adopted stellar parameters and their errors. The code estimates the masses using probability distribution functions (see Ramírez et al. (2013) for more details). The lithium abundances and estimated masses can be seen in Table A.6.

Our results have a very low scatter (0.04 dex) and the mean value A(Li)=+2.27 is compatible with measurements of similar metal-poor stars, such as those studied in Meléndez et al. (2010). As can be seen, the plateau is very well defined and the deviations are within our measurement errors.

Models of lithium depletion based on stellar evolution, such as Richard et al. (2005), predict that the least massive stars will be more depleted in Li. Meléndez et al. (2010) shows the existence of a correlation with the initial stellar mass using Richard et al. (2005) model predictions. The correlation found is especially good for the stars in the same metallicity range as this work. In Fig. 8 we show lithium abundances against the mass of the stars, based on Y2 isochrones. However, most of our targets have very similar masses which makes it very difficult to assess if there is any trend with mass.

For stars in the mass range we are working on (0.7 – 0.8 $M_{\odot}$), Fu et al. (2015) were able to reproduce the Spite plateau by invoking pre-stellar lithium depletion. In their model they take into consideration microscopic diffusion, overshooting, UV radiation photoevaporation and late accretion during the pre main sequence and main sequence phases. These effects are responsible for the lithium depletion in their model, which happens mainly in the pre main sequence phase and, to a lower extent, at the main sequence phase. Fu et al. (2015) calculated a A(Li)≈ 2.26, for stars with ages ranging from 10 to 12 Gyrs (see Fig 8 of Fu et al.), and reproduced the spite plateau over metallicities ranging from $-3.5 \leq$[M/H]$\leq -1.5$.

We also point out that stars with an even lower lithium abundance might have suffered effects from rotationally-induced mixing. Such effects have already been shown to deplete lithium in solar like stars (Carlos et al. 2016; Ryan et al. 2002) and might also be important to explain the lower lithium abundances found in some metal-poor stars, which might be the case of the two blue straggler stars found in our sample (see Sect. 7).

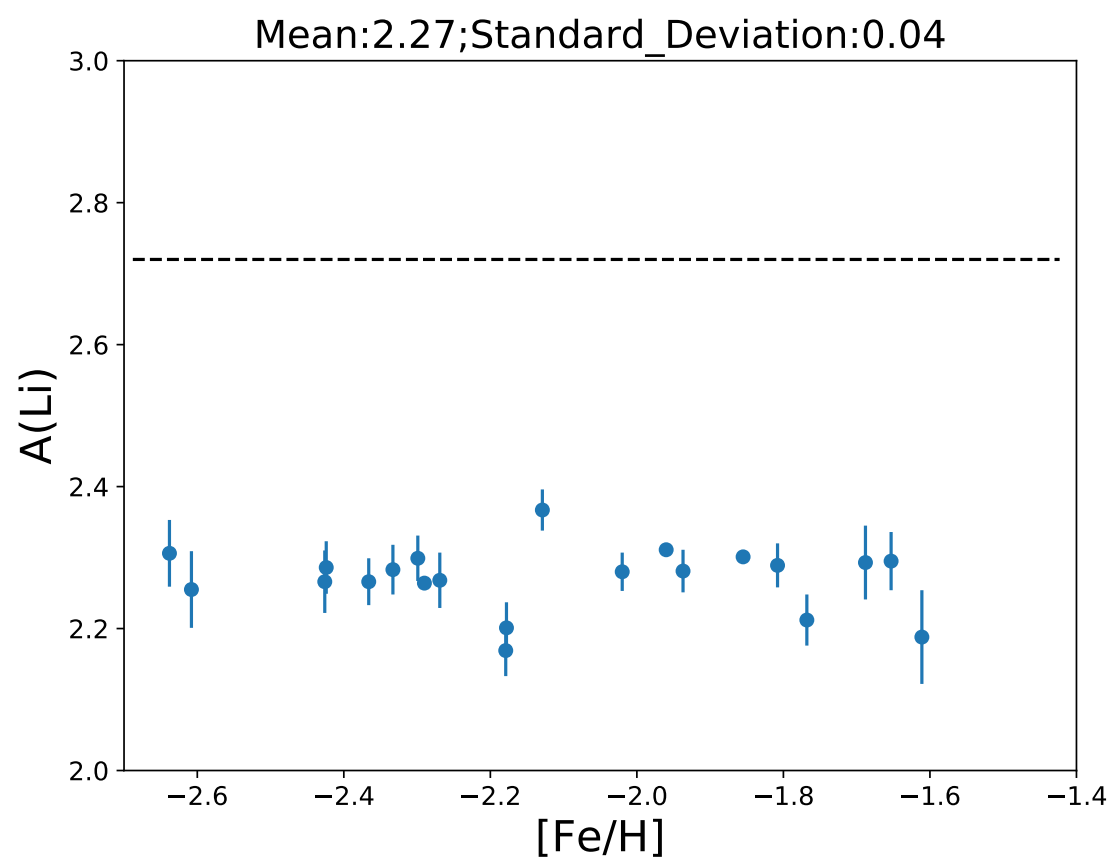

**Fig. 7.** A(Li) abundances of our sample. The dotted black line represent the Planck+BBN prediction (Coc et al. 2014b; Cyburt et al. 2016).

## 7. The blue straggler stars

Blue straggler stars (BSS) are main-sequence stars significantly bluer than the main-sequence turnoff population they belong to (Ryan et al. 2001). Due to the color difference from the regular main sequence stars, they are usually identified in Globular Clusters. Field BSS however are harder to identify because it is difficult to establish other main sequence stars with a common origin, to be used as standards in a color comparison. However, this identification is possible through other means, as employed by Santucci et al. (2015), who identified approximately 8000 BSS stars using color cuts, FWHM of the hydrogen spectral lines and stellar parameters.

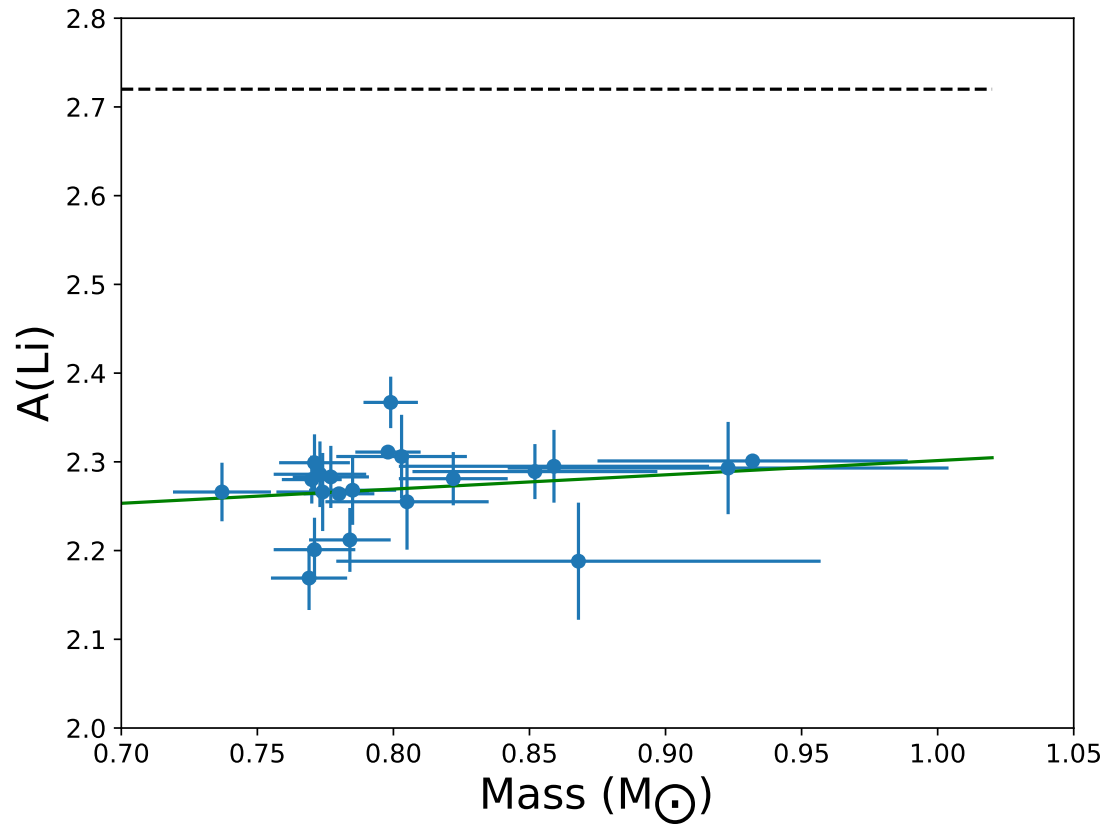

**Fig. 8.** Lithium vs. stellar mass. The green line represents a linear fit to the data. The black dotted line is the Planck and BBN prediction (Coc et al. 2014b; Cyburt et al. 2016).

Blue stragglers can also be identified via spectroscopy by using their Li or Be abundances. As showed by Ryan et al. (2001), halo ultra lithium-deficient stars can be BSS. We identified stars HD 340279 and G066-030 as blue stragglers based on their Li abundances, as was firstly done by Ryan et al. (2001). BS stars show much lower Li content when compared to stars of similar metallicity.

There are two possible scenarios for the formation of BS stars. In one of them the star is recipient of mass transfer from a more evolved AGB companion (the McCrea (1964) scenario), and in the second there was a collision with a companion. In both scenarios angular momentum is transfered to the BS star. This additional momentum can extend the convection zone, which is a possible explanation for part of the Li depletion. Blue straggler stars may also have enhanced s-process material, if it underwent mass transfer from an AGB companion. As stellar collisions will only occur in very dense environments, such as the core of globular clusters (Sills et al. 2009), it is more likely that the blue straggler stars found in the field have suffered mass transfer rather than collided.

We measured an upper limit to the Li abundance of star HD 340279 of A(Li)≤ 0.94 and A(Li)≤ 1.3 for G 66-30. Ryan et al. (2001) determined an upper limit of A(Li)≤ 1.39 for star HD 340279 and Boesgaard (2007) determined a conservative upper limit of A(Li)≤ 1.5 to G066-030. In both cases the stars are identified as BSS trough their ultra-deficient Li abundances compared to stars of similar effective temperature.

Boesgaard (2007) has also showed that G 66-30 is beryllium poor. They determined an upper limit of A(Be)< −1.0, which is below the expected value for Li normal stars, which also led to the conclusion that additional momentum has extended the convection zone and further depleted both elements.

Although there is a clear difference in Li abundance, not all other elements show such a clear difference, as can be seen in Figs. 2 to 6 (BS stars are the red objects). In Fig. 2 we see that the more metal-poor BSS may have a slight underabundance of $\alpha$-elements, while star G 66-30 is within the overall trends when the errors are taken into consideration.

The same effect happens when we look at the odd-Z light element Sc (Fig. 4), which is lower in HD 340279 when compared to stars of similar metallicity.

Star HD 340279 has another very puzzling peculiarity. Its vanadium abundance is lower than the abundances of all other stars. The calculated abundance is 0.6 dex lower when compared to the linear regression. It is not clear why there is such an underabundance, as the other BSS has a normal vanadium abundance and HD 340279 has normal abundances of the other iron peak elements, except for zinc. In Fig. 9 we show the spectra of stars HD 340279 and BD+26 2621, which have similar metallicities, around the 3952 V II line. We can see that the vanadium line of star BD+26 2621 is identified but in HD 340279 the line is barely visible. However, we caution the reader that the noise in our spectra is on the same level as the vanadium lines. Improved spectra are necessary to confirm this peculiarity.

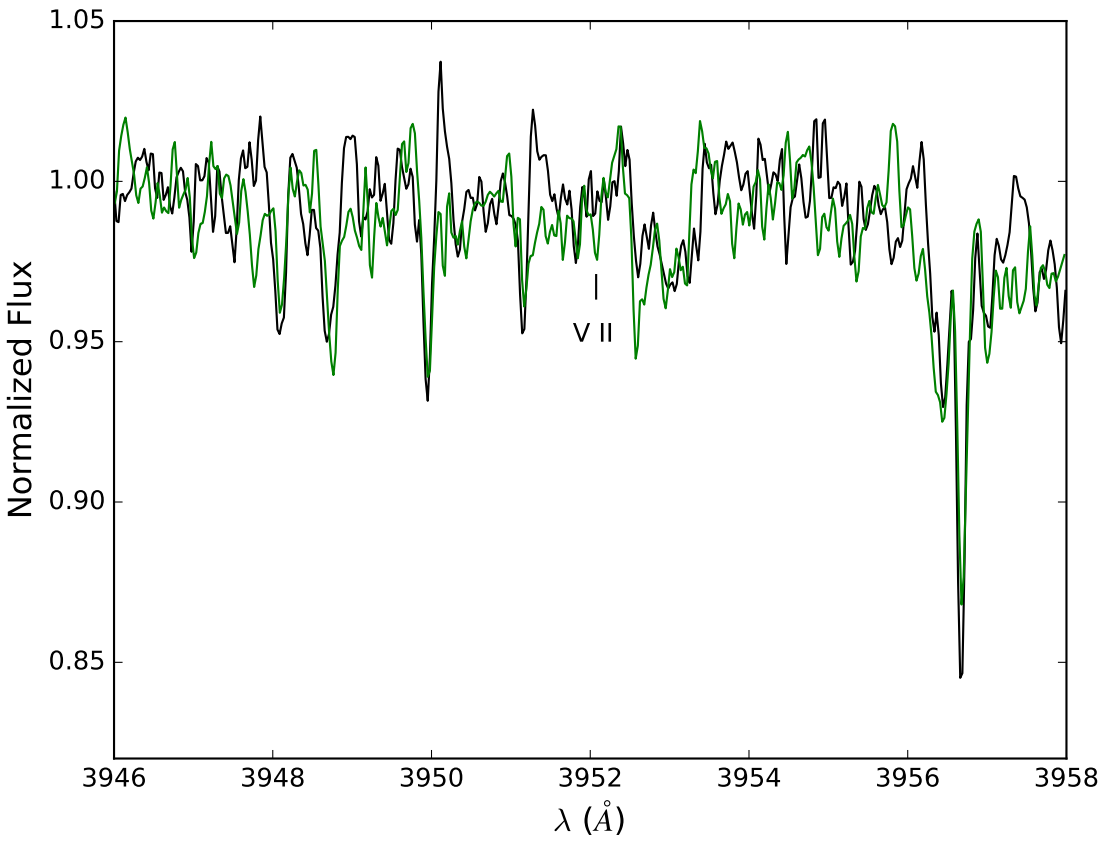

**Fig. 9.** Spectra around the V II line in 3952.02 for stars HD 340279 and BD+26 2621.

The BSS star HD 340279 might also be enhanced in zinc if compared exclusively with the rest of the stars of very similar metallicity, including the [Fe/H] = −2.58 star from Bonifacio et al. (2009). The zinc abundance of this star is also considerably higher than the linear regression predicts. However, if compared to the abundances of the more metal-poor stars of Bonifacio et al. (2009), one could attribute the enhancement to cosmic scatter.

Considering the possibility of Zn enhancement in HD 340279, we can discuss its cause in light of the nucleosynthetic processes that a BSS star undergoes. While studying nucleosynthesis in POP III stars Heger & Woosley (2002) and Umeda & Nomoto (2002), first proposed the possibility of Zn being produced by the s-process. Although zinc production in AGB stars is not large enough to cause comprehensive changes to the GCE (e.g., Karakas et al. 2009), it is possible that a star that underwent mass transfer from an AGB can show enhanced Zn. Under that assumption the excess of zinc in BSS could be another tool to estimate the mass of the AGB progenitor, as production of Zn in the AGB phase is more important in intermediate $M \geq 3M_{\odot}$ AGB stars. Zinc is at the beginning of the s-process chain but overall production is low, on the order of [Zn/Fe] < 0.3. The exception is in intermediate mass AGB stars where [Zn/Fe] ∼ 0.5.

In order to confirm if there is an excess of Zn in this star or if that is just an effect of cosmic scatter, it is of extreme importance to obtain more Zn abundances of BSS stars, providing tools to constrain a possible enhancement in the BSS process, or to exclude the possibility. Strontium, yttrium and barium are very enhanced in HD 340279. It indicates a very big influence of s-process nucleosynthesis in this star. G66-030 on the other hand, does not seem to have an enhanced s-process and the abundance is low when compared to the other objects.

Unfortunately the spectroscopic works on BSS that we found are usually focused on one or just a few elements, as for example in Ferraro et al. (2006, 2016). We emphasize the importance of more comprehensive studies on the abundance patterns of BSS.

## 8. Conclusions

Previous studies of metal-poor stars in the halo are mainly focused on the most metal-poor end, [Fe/H]≤ −2.5 (e.g., Ryan et al. 1996; Norris et al. 2001; Cayrel et al. 2004; Yong et al. 2013; Norris et al. 2013). Our study, on the other hand, has metallicities [Fe/H]≥ −2.7. Our sample was chosen to obtain good spectra of similar stars, allowing us to seek precise abundances.

The differential analysis technique allow us to greatly decrease the data scatter, and also the errors of the differential chemical abundances. The well-defined trends we observe in our results are, for most elements, compatible with what has been previously found at lower precision.

The small scatter in our data set shows that the chemical evolution was, overall, very homogeneous. Among $\alpha$-elements, our data for Ca is in excellent agreement with the GCE model while we found ∼ 0.2 dex offset for Mg, in concordance with the NLTE study of Zhao et al. (2016). We do not see any indication of different populations in $\alpha$-elements, as found by Nissen & Schuster (2010), below their metallicity threshold. If the stars in their sample were acreted from satellite galaxies, the fact that we do not see these populations in our sample can indicate that the main accretion events started to take place only after SNe Ia already had time to increase the overall metallicity to [Fe/H]≈ −1.5. This is supported by the fact that the separation between the two populations is mostly seen in the more metal-rich end, and seems to become homogeneous at the metal-poor end of their sample.

The very good homogeneity we observe in the $\alpha$-elements is also seen in odd-light elements and iron peak elements. The Co and Zn trends at very low metallicities also suggests an inhomogeneous enrichment with hypernovae.

Our differential LTE analysis shows a very good reliability even when compared to a full NLTE analysis. As shown, the results we obtain are very similar to the NLTE analysis of Zhao et al. (2016), and the comparison with the K15 GCE model they performed are very similar to those presented in this work. Although there is a very good agreement, in some cases the NLTE approach decreases the discrepancy between model and observations, as is the case of scandium, where there is a mean difference of ≈ 0.1 dex between ours and their abundances. In the case of calcium, however, the results are very much alike and both NLTE and LTE have an impressive agreement with the K15 model. It is also important to stress that the values we compared to the K15 model are our mean values, and also the difference between data and model is based on our own measurements. Zhao et al. (2016) studied a broader metallicity range and thus is internally consistent, unlike our comparisons that are less homogeneous as employed data from other works, however overall there seems to be a good connection between our work and lower and higher metallicities.

LTE calculations of Cr II are much more reliable than Cr I. As shown by Bergemann & Cescutti (2010), Cr I suffers from strong NLTE effects due to the over-ionization from the low-excitation odd Cr I levels, which is more severe in more metal-poor stars, explaining the positive slope seen in the Cr I data, while Cr II NLTE effects on abundances are negligible for dwarfs. Taking Cr II as the indicative of [Cr/Fe], we see a good agreement between GCE predictions and observed abundances for the entire metallicity range ($-3.6 \leq$ [Fe/H] $\leq -0.4$), indicative of the good understanding of Cr nucleosynthesis.

The bigger offsets between observational data and GCE predictions are seen for elements with an already known deviation, such as Co or Mn, which are in agreement with other observational works (e.g, Cayrel et al. 2004; Bonifacio et al. 2009) but somehow still far from GCE models. For these elements there might be important NLTE effects that are not being taken into consideration in this work, such as for Mn (Bergemann & Gehren 2008). This disagreement between observation and GCE predictions is not seen in Kobayashi et al. (2006), for example, as the absolute [X/Fe] values for most elements agree with observations, because the observations are spread due to results from different authors, making the comparison samples inhomogeneous. This is also shown in Fig. 3, by the big dispersion we found using data from the SAGA database. The Co and Zn trends also suggests an inhomogeneous enrichment with hypernovae.

Although AGB stars do not produce enough zinc to influence Galactic chemical evolution, the fact that the BS star HD 340279 may be enhanced in zinc, suggests that this could be an important tracer of the AGB progenitor masses of BS stars. Also, further observations of neutron-capture elements for this star might be beneficial to constrain the origin of Zn in this object.

It is also worth commenting on the analysis of star WISE J072543.88-235119.7, a high proper motion star crossing the Galactic plane with a bound retrograde orbit. Scholz et al. (2015) used a spectrum of lower resolution and S/N than ours, finding stellar parameters that are roughly consistent with the parameters found in our work. They suggested that this star might be a good target for follow-up high-resolution spectroscopy. In our results we did not find any distinctive chemical peculiarity in this star. The abundance pattern seems to be in good agreement with the remaining of our sample. Spite et al. (2015) performed a spectroscopic analysis of this star, calculating the effective temperature using a different method (H$\alpha$ fitting), resulting in a different set of stellar parameters. The differences in the stellar parameters translated into the abundance differences between our work and theirs. We verified this by recalculating the abundances using our equivalent widths and their stellar parameters, showing a mean difference of 0.08 dex in A(X), which can be easily explained by the measurement errors of both works. Thus, the differences between our measurements arise from the difference in the stellar parameters. They conclude through the Li abundance that despite the extreme kinematics the star might have formed in situ, which is in agreement with our findings, as it has an abundance pattern resembling that of our own galaxy.

Here we used a line-by-line differential work to better constrain the chemical evolution of the Galaxy in a metallicity range that does not have many high precision works. Our abundances, along with the data from works in other metallicity ranges, allow us to do a comprehensive comparison of observational data of stars with similar atmospheric parameters, to the K15 Galactic evolution model. Our careful analysis yields precise and accurate data, which have small errors and low scatter, being thus important to better constrain future developments of GCE models. Finally, we encourage NLTE calculations in further works, in particular for the elements Na, Al, Si, Sc, and Ba.

*Acknowledgements.* This work was based on observations collected at the European Organisation for Astronomical Research in the Southern Hemisphere under ESO program 095.D-0504(A). We would like to thank the referee for his/her valuable inputs. HR thanks a CAPES fellowship and CAPES PDSE program (88881.132145/2016-01). JM thanks support by FAPESP (2012/24392-2 and 2014/18100-4) and CNPq (Productivity Fellowship). V.M.P. acknowledges partial support for this work from the National Science Foundation under Grant No. PHY-1430152 (JINA Center for the Evolution of the Elements).

## Appendix A: Stellar parameters and chemical abundances.

**Table A.1.** Stellar parameters for each star. The standard stars are in bold and with *. Superscript numbers 1 and 2 represent the samples compared to the standard stars HD 338529 and CD-48 2445, respectively.

| Star | $T_{\rm eff}$ K | $\sigma T_{\rm eff}$ K | log $g$ dex | $\sigma$log $g$ dex | $v_T$ kms$^{-1}$ | $\sigma v_T$ kms$^{-1}$ | [Fe/H] dex | $\sigma$[Fe/H] dex |
|---|---|---|---|---|---|---|---|---|
| BD+20 3603[1] | 6229 | 50 | 4.09 | 0.08 | 1.29 | 0.04 | −2.179 | 0.042 |
| BD+24 1676[1] | 6438 | 63 | 4.13 | 0.10 | 1.54 | 0.04 | −2.426 | 0.049 |
| BD+26 2621[1] | 6470 | 81 | 4.51 | 0.12 | 1.34 | 0.05 | −2.608 | 0.063 |
| BD-04 3208[1] | 6433 | 49 | 4.11 | 0.07 | 1.55 | 0.03 | −2.333 | 0.037 |
| BD-13 3442[1] | 6569 | 73 | 4.36 | 0.12 | 1.67 | 0.05 | −2.638 | 0.054 |
| CD-71 1234[1] | 6421 | 53 | 4.31 | 0.09 | 1.46 | 0.03 | −2.424 | 0.040 |
| BPS CS 22943-0095[1] | 6414 | 46 | 4.27 | 0.07 | 1.42 | 0.04 | −2.299 | 0.036 |
| G 126-52[1] | 6462 | 57 | 4.28 | 0.09 | 1.47 | 0.04 | −2.269 | 0.043 |
| **HD 338529*** | 6426 | 50 | 4.09 | 0.03 | 1.50 | 0.05 | −2.290 | 0.050 |
| HD 340279[1] | 6493 | 70 | 4.52 | 0.09 | 1.29 | 0.05 | −2.561 | 0.055 |
| LP 894-1[1] | 6378 | 53 | 4.26 | 0.09 | 1.37 | 0.03 | −2.178 | 0.041 |
| WISE J072543.88-235119.7[2] | 6160 | 45 | 4.42 | 0.09 | 1.30 | 0.04 | −2.366 | 0.038 |
| BD+01 3597[2] | 6435 | 44 | 4.04 | 0.07 | 1.57 | 0.03 | −1.937 | 0.035 |
| BD+02 4651[2] | 6241 | 43 | 3.89 | 0.09 | 1.49 | 0.03 | −1.808 | 0.036 |
| **CD-48 2445*** | 6453 | 50 | 4.23 | 0.03 | 1.50 | 0.05 | −1.960 | 0.050 |
| G 66-30[2] | 6638 | 47 | 4.36 | 0.09 | 1.52 | 0.05 | −1.473 | 0.038 |
| G 126-62[2] | 6145 | 90 | 3.91 | 0.18 | 1.13 | 0.15 | −1.611 | 0.097 |
| HD 59392[2] | 6056 | 72 | 3.72 | 0.11 | 1.28 | 0.10 | −1.688 | 0.075 |
| HD 74000[2] | 6341 | 39 | 4.19 | 0.06 | 1.46 | 0.03 | −2.020 | 0.031 |
| HD 84937[2] | 6513 | 44 | 4.17 | 0.06 | 1.61 | 0.04 | −2.129 | 0.032 |
| HD 108177[2] | 6107 | 50 | 4.04 | 0.06 | 1.17 | 0.07 | −1.768 | 0.050 |
| HD 110621[2] | 6182 | 56 | 3.9 | 0.11 | 1.34 | 0.07 | −1.653 | 0.054 |
| HD 122196[2] | 6052 | 52 | 3.66 | 0.07 | 1.44 | 0.05 | −1.855 | 0.048 |

**Table A.2.** Abundances of the $\alpha$-elements.

| Star | [Mg/Fe] | $\sigma$_[Mg/Fe] | [Si/Fe] | $\sigma$_[Si/Fe] | [Ca/Fe] | $\sigma$_[Ca/Fe] | [Ti/Fe] | $\sigma$_[Ti/Fe] |
|---|---|---|---|---|---|---|---|---|
| BD +203603 | 0.343 | 0.062 | 0.203 | 0.081 | 0.369 | 0.056 | 0.358 | 0.06 |
| BD +241676 | 0.343 | 0.072 | 0.125 | 0.08 | 0.414 | 0.063 | 0.482 | 0.069 |
| BD +262621 | 0.319 | 0.088 | 0.116 | 0.101 | 0.382 | 0.083 | 0.426 | 0.089 |
| BD -043208 | 0.361 | 0.055 | 0.153 | 0.064 | 0.455 | 0.051 | 0.507 | 0.056 |
| BD -133442 | 0.401 | 0.084 | 0.049 | 0.081 | 0.439 | 0.07 | 0.574 | 0.075 |
| CD-71 1234 | 0.272 | 0.064 | 0.165 | 0.069 | 0.354 | 0.054 | 0.388 | 0.059 |
| BPS CS 22943-0095 | 0.397 | 0.052 | 0.23 | 0.064 | 0.429 | 0.048 | 0.491 | 0.055 |
| G 126-52 | 0.25 | 0.063 | 0.09 | 0.073 | 0.358 | 0.058 | 0.443 | 0.064 |
| HD 338529 | 0.386 | 0 | 0.232 | 0 | 0.442 | 0 | 0.465 | 0 |
| HD 340279 | 0.202 | 0.076 | 0.022 | 0.084 | 0.289 | 0.071 | 0.358 | 0.079 |
| LP 894-1 | 0.253 | 0.062 | 0.135 | 0.075 | 0.337 | 0.055 | 0.389 | 0.061 |
| WISE J072543.88-235119.7 | 0.314 | 0.059 | 0.226 | 0.072 | 0.28 | 0.049 | 0.326 | 0.059 |
| BD +013597 | 0.3 | 0.055 | 0.164 | 0.063 | 0.403 | 0.046 | 0.467 | 0.058 |
| BD +024651 | 0.33 | 0.059 | 0.184 | 0.067 | 0.395 | 0.048 | 0.391 | 0.057 |
| CD-48 2445 | 0.249 | 0 | 0.142 | 0 | 0.346 | 0 | 0.434 | 0 |
| G 66-30 | 0.212 | 0.087 | 0.114 | 0.069 | 0.278 | 0.051 | 0.319 | 0.058 |
| G 126-62 | 0.26 | 0.152 | 0.08 | 0.164 | 0.379 | 0.129 | 0.344 | 0.125 |
| HD 59392 | 0.315 | 0.116 | 0.16 | 0.123 | 0.365 | 0.098 | 0.362 | 0.096 |
| HD 74000 | 0.34 | 0.054 | 0.228 | 0.058 | 0.347 | 0.041 | 0.376 | 0.044 |
| HD 84937 | 0.288 | 0.049 | 0.165 | 0.057 | 0.375 | 0.043 | 0.475 | 0.052 |
| HD 108177 | 0.38 | 0.082 | 0.204 | 0.086 | 0.361 | 0.066 | 0.353 | 0.062 |
| HD 110621 | 0.392 | 0.086 | 0.249 | 0.096 | 0.395 | 0.072 | 0.389 | 0.072 |
| HD 122196 | 0.24 | 0.03 | 0.128 | 0.068 | 0.286 | 0.024 | 0.275 | 0.036 |

**Table A.3.** Abundances of the light odd Z elements.

| Star | $[Na/Fe]_{NLTE}$ | $\sigma$_[Na/Fe] | $[Al_{NLTE}/Fe]$ | $\sigma$_$[Al_{NLTE}/Fe]$ | [Sc/Fe] | $\sigma$_[Sc/Fe] |
|---|---|---|---|---|---|---|
| BD +203603 | −0.276 | 0.089 | −0.154 | 0.083 | 0.252 | 0.060 |
| BD +241676 | −0.123 | 0.124 | −0.095 | 0.072 | 0.360 | 0.071 |
| BD +262621 | −0.307 | 0.091 | −0.160 | 0.096 | 0.238 | 0.112 |
| BD -043208 | −0.216 | 0.054 | −0.092 | 0.059 | 0.354 | 0.057 |
| BD -133442 | −0.103 | 0.076 | −0.076 | 0.077 | 0.493 | 0.085 |
| CD-71 1234 | −0.243 | 0.072 | −0.221 | 0.064 | 0.256 | 0.066 |
| BPS CS 22943-0095 | −0.043 | 0.055 | −0.082 | 0.063 | 0.339 | 0.053 |
| G 126-52 | −0.159 | 0.067 | −0.224 | 0.067 | 0.316 | 0.065 |
| HD 338529 | −0.153 | 0 | −0.093 | 0 | 0.345 | 0 |
| HD 340279 | −0.255 | 0.158 | −0.131 | 0.077 | 0.040 | 0.150 |
| LP 894-1 | −0.344 | 0.090 | −0.233 | 0.062 | 0.245 | 0.062 |
| WISE J072543.88-235119.7 | −0.388 | 0.061 | −0.137 | 0.061 | 0.207 | 0.063 |
| BD +013597 | −0.255 | 0.062 | −0.108 | 0.054 | 0.369 | 0.052 |
| BD +024651 | −0.215 | 0.091 | −0.174 | 0.115 | 0.332 | 0.057 |
| CD-48 2445 | −0.516 | 0 | −0.212 | 0 | 0.319 | 0 |
| G 66-30 | −0.279 | 0.092 | −0.242 | 0.060 | 0.264 | 0.057 |
| G 126-62 | −0.243 | 0.183 | −0.162 | 0.181 | 0.351 | 0.141 |
| HD 59392 | −0.203 | 0.185 | −0.090 | 0.225 | 0.331 | 0.116 |
| HD 74000 | 0.061 | 0.083 | 0.094 | 0.079 | 0.242 | 0.047 |
| HD 84937 | −0.234 | 0.125 | −0.112 | 0.057 | 0.332 | 0.054 |
| HD 108177 | −0.074 | 0.114 | −0.053 | 0.134 | 0.264 | 0.069 |
| HD 110621 | −0.160 | 0.120 | 0.006 | 0.167 | 0.374 | 0.085 |
| HD 122196 | −0.257 | 0.058 | −0.183 | 0.149 | 0.205 | 0.034 |

**Table A.4.** Abundances of the iron peak elements.

| Star | [V/Fe] | σ_[V/Fe] | [CrI/Fe] | σ_[CrI/Fe] | [CrII/Fe] | σ_[CrII/Fe] | [Mn/Fe] | σ_[Mn/Fe] | [Co/Fe] | σ_[Co/Fe] | [Ni/Fe] | σ_[Ni/Fe] | [Zn/Fe] | σ_[Zn/Fe] |
|---|---|---|---|---|---|---|---|---|---|---|---|---|---|---|
| BD +203603 | 0.125 | 0.060 | −0.138 | 0.078 | 0.087 | 0.090 | −0.525 | 0.074 | 0.114 | 0.076 | −0.021 | 0.086 | 0.259 | 0.051 |
| BD +241676 | 0.227 | 0.074 | −0.143 | 0.085 | 0.153 | 0.074 | −0.561 | 0.083 | 0.287 | 0.093 | 0.005 | 0.096 | 0.386 | 0.059 |
| BD +262621 | 0.355 | 0.085 | −0.190 | 0.102 | 0.139 | 0.087 | −0.653 | 0.126 | 0.266 | 0.113 | −0.040 | 0.118 | 0.454 | 0.075 |
| BD -043208 | 0.218 | 0.052 | −0.143 | 0.068 | 0.059 | 0.059 | −0.604 | 0.065 | 0.194 | 0.070 | −0.090 | 0.078 | 0.437 | 0.047 |
| BD -133442 | 0.560 | 0.134 | −0.111 | 0.087 | 0.204 | 0.077 | −0.567 | 0.094 | 0.306 | 0.094 | −0.055 | 0.099 | 0.224 | 0.066 |
| CD-71 1234 | 0.220 | 0.057 | −0.174 | 0.071 | 0.042 | 0.072 | −0.597 | 0.068 | 0.216 | 0.075 | −0.059 | 0.080 | 0.367 | 0.050 |
| BPS CS 22943-0095 | 0.209 | 0.062 | −0.164 | 0.060 | 0.113 | 0.057 | −0.579 | 0.060 | 0.199 | 0.072 | −0.003 | 0.074 | 0.381 | 0.044 |
| G 126-52 | 0.366 | 0.215 | −0.112 | 0.081 | 0.182 | 0.098 | −0.441 | 0.075 | 0.174 | 0.085 | −0.005 | 0.088 | 0.195 | 0.053 |
| HD 338529 | 0.228 | 0 | −0.159 | 0 | 0.127 | 0 | −0.531 | 0 | 0.177 | 0 | −0.039 | 0 | 0.433 | 0 |
| HD 340279 | −0.260 | 0.072 | −0.114 | 0.087 | 0.223 | 0.133 | −0.707 | 0.094 | 0.268 | 0.105 | −0.120 | 0.103 | 0.551 | 0.064 |
| LP 894-1 | 0.237 | 0.167 | −0.145 | 0.078 | 0.086 | 0.081 | −0.609 | 0.068 | 0.095 | 0.076 | −0.089 | 0.083 | 0.364 | 0.051 |
| WISE J072543.88-235119.7 | 0.044 | 0.054 | 0.056 | 0.072 | 0.089 | 0.060 | −0.495 | 0.078 | 0.232 | 0.066 | −0.015 | 0.076 | 0.228 | 0.045 |
| BD +013597 | 0.131 | 0.058 | 0.214 | 0.079 | 0.113 | 0.055 | −0.408 | 0.089 | 0.110 | 0.067 | −0.013 | 0.073 | 0.328 | 0.043 |
| BD +024651 | 0.135 | 0.052 | 0.181 | 0.063 | 0.126 | 0.074 | −0.418 | 0.057 | 0.065 | 0.069 | −0.053 | 0.074 | 0.245 | 0.044 |
| CD-48 2445 | 0.147 | 0 | 0.181 | 0 | 0.163 | 0 | −0.456 | 0 | 0.131 | 0 | −0.043 | 0 | 0.251 | 0 |
| G 66-30 | 0.149 | 0.052 | 0.146 | 0.072 | 0.076 | 0.070 | −0.369 | 0.073 | 0.043 | 0.073 | −0.046 | 0.080 | 0.243 | 0.047 |
| G 126-62 | 0.098 | 0.124 | 0.180 | 0.170 | 0.216 | 0.142 | −0.404 | 0.144 | −0.078 | 0.157 | −0.053 | 0.189 | 0.096 | 0.112 |
| HD 59392 | 0.060 | 0.102 | 0.188 | 0.142 | 0.187 | 0.115 | −0.402 | 0.108 | 0 | 0.125 | −0.047 | 0.146 | 0.248 | 0.086 |
| HD 74000 | 0.129 | 0.041 | 0.176 | 0.053 | 0.099 | 0.069 | −0.437 | 0.054 | 0.122 | 0.055 | −0.014 | 0.064 | 0.328 | 0.037 |
| HD 84937 | 0.243 | 0.147 | 0.143 | 0.058 | 0.126 | 0.047 | −0.432 | 0.076 | 0.175 | 0.060 | −0.018 | 0.067 | 0.294 | 0.040 |
| HD 108177 | 0.092 | 0.061 | 0.170 | 0.095 | 0.095 | 0.074 | −0.426 | 0.074 | 0.017 | 0.088 | −0.026 | 0.098 | 0.267 | 0.057 |
| HD 110621 | 0.160 | 0.071 | 0.195 | 0.106 | 0.112 | 0.085 | −0.357 | 0.083 | 0.069 | 0.099 | 0.016 | 0.110 | 0.230 | 0.064 |
| HD122196 | −0.004 | 0.033 | 0.123 | 0.051 | 0.112 | 0.057 | −0.467 | 0.028 | −0.053 | 0.041 | −0.103 | 0.048 | 0.177 | 0.020 |

**Table A.5.** Abundances of the neutron-capture elements.

| Star | [Sr/Fe] | σ_[Sr/Fe] | [Y/Fe] | σ_[Y/Fe] | [Zr/Fe] | σ_[Zr/Fe] | [Ba/Fe] | σ_[Ba/Fe] |
|---|---|---|---|---|---|---|---|---|
| BD +203603 | 0.026 | 0.065 | −0.235 | 0.064 | 0.376 | 0.091 | −0.316 | 0.067 |
| BD +241676 | 0.094 | 0.080 | −0.009 | 0.084 | 0.648 | 0.067 | −0.268 | 0.069 |
| BD +262621 | −0.287 | 0.091 | −0.181 | 0.109 | 0.410 | 0.087 | −0.590 | 0.091 |
| BD -043208 | 0.179 | 0.057 | −0.013 | 0.057 | 0.532 | 0.082 | −0.212 | 0.066 |
| BD -133442 | 0.184 | 0.086 | 0.124 | 0.090 | 0.636 | 0.125 | −0.532 | 0.079 |
| CD-71 1234 | −0.616 | 0.068 | −0.652 | 0.068 | 0.617 | 0.268 | −0.441 | 0.068 |
| BPS CS 22943-0095 | 0.321 | 0.058 | 0.119 | 0.064 | 0.711 | 0.094 | −0.075 | 0.074 |
| G 126-52 | −0.018 | 0.07 | −0.032 | 0.089 | 0.840 | 0.060 | −0.134 | 0.062 |
| HD 338529 | 0.129 | 0 | −0.040 | 0 | 0.559 | 0 | −0.058 | 0 |
| HD 340279 | 0.359 | 0.103 | 0.321 | 0.076 | 0.623 | 0.095 | 0.394 | 0.078 |
| LP 894-1 | 0.089 | 0.067 | −0.137 | 0.071 | 0.567 | 0.088 | 0.024 | 0.067 |
| WISE J072543.88-235119.7 | −0.006 | 0.058 | −0.022 | 0.061 | 0.473 | 0.071 | 0.175 | 0.060 |
| BD +013597 | 0.162 | 0.052 | 0.023 | 0.059 | 0.527 | 0.055 | 0.014 | 0.005 |
| BD +024651 | 0.139 | 0.057 | −0.077 | 0.061 | 0.459 | 0.081 | 0.206 | 0.061 |
| CD-48 2445 | 0.207 | 0 | 0.069 | 0 | 0.673 | 0 | 0.334 | 0 |
| G 66-30 | 0.109 | 0.070 | −0.153 | 0.059 | 0.496 | 0.083 | 0.140 | 0.066 |
| G 126-62 | 0.134 | 0.151 | −0.141 | 0.141 | 0.444 | 0.155 | 0.137 | 0.153 |
| HD 59392 | 0.242 | 0.109 | 0.043 | 0.112 | 0.576 | 0.105 | 0.380 | 0.129 |
| HD 74000 | 0.283 | 0.048 | 0.028 | 0.049 | 0.592 | 0.075 | 0.221 | 0.061 |
| HD 84937 | 0.075 | 0.062 | −0.027 | 0.052 | 0.552 | 0.106 | −0.043 | 0.070 |
| HD 108177 | 0.171 | 0.076 | −0.083 | 0.077 | 0.484 | 0.119 | 0.043 | 0.095 |
| HD 110621 | 0.281 | 0.085 | 0.011 | 0.092 | 0.560 | 0.101 | 0.272 | 0.093 |
| HD122196 | −0.007 | 0.046 | −0.298 | 0.036 | 0.309 | 0.079 | 0.025 | 0.042 |

**Table A.6.** Lithium abundances and mass estimates for our stars. We note that the two blue straggler stars are not included in this table.

| Star | A(Li) dex | σA(Li) dex | Mass $M_\odot$ |
|---|---|---|---|
| BD+20 3603 | 2.169 | 0.036 | 0.769 |
| BD+24 1676 | 2.266 | 0.044 | 0.774 |
| BD+26 2621 | 2.255 | 0.054 | 0.805 |
| BD-04 3208 | 2.283 | 0.035 | 0.777 |
| BD-13 3442 | 2.306 | 0.047 | 0.803 |
| CD-71 1234 | 2.286 | 0.037 | 0.773 |
| BPS CS 22943-0095 | 2.299 | 0.032 | 0.771 |
| G 126-52 | 2.268 | 0.039 | 0.785 |
| HD 338529 | 2.264 | 0.035 | 0.780 |
| LP 894-1 | 2.201 | 0.036 | 0.771 |
| WISE J072543.88-235119.7 | 2.266 | 0.033 | 0.737 |
| BD+01 3597 | 2.281 | 0.030 | 0.822 |
| BD+02 4651 | 2.289 | 0.031 | 0.852 |
| CD-48 2445 | 2.311 | 0.032 | 0.798 |
| G 126-62 | 2.188 | 0.066 | 0.868 |
| HD 59392 | 2.293 | 0.052 | 0.923 |
| HD 74000 | 2.280 | 0.027 | 0.770 |
| HD 84937 | 2.367 | 0.029 | 0.799 |
| HD 108177 | 2.212 | 0.036 | 0.784 |
| HD 110621 | 2.295 | 0.041 | 0.859 |
| HD 122196 | 2.301 | 0.020 | 0.932 |

## Appendix B: Linelist

**Table B.1.** Linelist used for the abundances determinations. The linelist is formatted to be used with the radiative transfer code MOOG (Sneden 1973), and also include the hyperfine splitting, indicated by the negative wavelengths.

| Wavelength (Å) | Species | EP (eV) | log($gf$) (dex) |
|---|---|---|---|
| 3902.95 | 26.0 | 1.56 | −0.47 |
| 3906.48 | 26.0 | 0.11 | −2.24 |
| 3917.18 | 26.0 | 0.99 | −2.16 |
| 3920.26 | 26.0 | 0.12 | −1.75 |
| 3922.91 | 26.0 | 0.05 | −1.65 |
| 3927.92 | 26.0 | 0.11 | −1.52 |
| 3930.30 | 26.0 | 0.09 | −1.49 |
| 3940.88 | 26.0 | 0.96 | −2.60 |
| 3949.95 | 26.0 | 2.18 | −1.25 |
| 3977.74 | 26.0 | 2.20 | −1.12 |
| 3997.39 | 26.0 | 2.73 | −0.48 |
| 3998.05 | 26.0 | 2.69 | −0.91 |
| 4005.24 | 26.0 | 1.56 | −0.61 |
| 4021.87 | 26.0 | 2.76 | −0.73 |
| 4045.81 | 26.0 | 1.49 | 0.28 |
| 4063.59 | 26.0 | 1.56 | 0.06 |
| 4071.74 | 26.0 | 1.61 | −0.02 |
| 4118.55 | 26.0 | 3.57 | 0.22 |
| 4134.68 | 26.0 | 2.83 | −0.65 |
| 4143.42 | 26.0 | 3.05 | −0.20 |
| 4143.87 | 26.0 | 1.56 | −0.51 |
| 4147.67 | 26.0 | 1.49 | −2.10 |
| 4154.50 | 26.0 | 2.83 | −0.69 |
| 4154.81 | 26.0 | 3.37 | −0.40 |
| 4156.80 | 26.0 | 2.83 | −0.81 |
| 4175.64 | 26.0 | 2.85 | −0.83 |
| 4181.76 | 26.0 | 2.83 | −0.37 |
| 4187.04 | 26.0 | 2.45 | −0.55 |
| 4187.80 | 26.0 | 2.43 | −0.55 |
| 4191.43 | 26.0 | 2.47 | −0.67 |
| 4199.10 | 26.0 | 3.05 | 0.16 |
| 4202.03 | 26.0 | 1.49 | −0.71 |
| 4216.18 | 26.0 | 0.00 | −3.36 |
| 4222.21 | 26.0 | 2.45 | −0.97 |
| 4227.43 | 26.0 | 3.33 | 0.27 |
| 4233.60 | 26.0 | 2.48 | −0.60 |
| 4238.81 | 26.0 | 3.40 | −0.23 |
| 4250.12 | 26.0 | 2.47 | −0.41 |
| 4250.79 | 26.0 | 1.56 | −0.71 |
| 4260.47 | 26.0 | 2.40 | 0.11 |
| 4271.15 | 26.0 | 2.45 | −0.35 |
| 4271.76 | 26.0 | 1.49 | −0.16 |
| 4282.40 | 26.0 | 2.18 | −0.78 |
| 4375.93 | 26.0 | 0.00 | −3.03 |
| 4383.55 | 26.0 | 1.49 | 0.20 |
| 4404.75 | 26.0 | 1.56 | −0.14 |
| 4427.31 | 26.0 | 0.05 | −2.92 |
| 4442.34 | 26.0 | 2.20 | −1.26 |
| 4459.12 | 26.0 | 2.18 | −1.28 |
| 4461.65 | 26.0 | 0.09 | −3.21 |
| 4466.55 | 26.0 | 2.83 | −0.60 |
| 4494.56 | 26.0 | 2.20 | −1.14 |
| 4871.32 | 26.0 | 2.87 | −0.36 |
| 4872.14 | 26.0 | 2.88 | −0.57 |
| 4890.76 | 26.0 | 2.88 | −0.39 |
| 4891.49 | 26.0 | 2.85 | −0.11 |
| 4918.99 | 26.0 | 2.87 | −0.34 |
| 4920.50 | 26.0 | 2.83 | 0.07 |

**Table B.1.** Continued

| Wavelength (Å) | Species | EP (eV) | log($gf$) (dex) |
|---|---|---|---|
| 4957.30 | 26.0 | 2.85 | −0.41 |
| 5049.82 | 26.0 | 2.28 | −1.36 |
| 5133.69 | 26.0 | 4.18 | 0.14 |
| 5139.25 | 26.0 | 3.00 | −0.74 |
| 5139.46 | 26.0 | 2.94 | −0.51 |
| 5162.27 | 26.0 | 4.18 | 0.02 |
| 5171.60 | 26.0 | 1.49 | −1.79 |
| 5191.46 | 26.0 | 3.04 | −0.55 |
| 5194.94 | 26.0 | 1.56 | −2.09 |
| 5216.27 | 26.0 | 1.61 | −2.15 |
| 5226.86 | 26.0 | 3.04 | −0.56 |
| 5227.19 | 26.0 | 1.56 | −1.23 |
| 5232.94 | 26.0 | 2.94 | −0.06 |
| 5266.56 | 26.0 | 3.00 | −0.39 |
| 5328.04 | 26.0 | 0.92 | −1.47 |
| 5328.53 | 26.0 | 1.56 | −1.85 |
| 5369.96 | 26.0 | 4.37 | 0.54 |
| 5383.37 | 26.0 | 4.31 | 0.65 |
| 5397.13 | 26.0 | 0.92 | −1.99 |
| 5405.78 | 26.0 | 0.99 | −1.84 |
| 5415.20 | 26.0 | 4.39 | 0.64 |
| 5424.07 | 26.0 | 4.32 | 0.52 |
| 5429.70 | 26.0 | 0.96 | −1.88 |
| 5434.52 | 26.0 | 1.01 | −2.12 |
| 5455.61 | 26.0 | 1.01 | −2.09 |
| 5497.52 | 26.0 | 1.01 | −2.85 |
| 5506.78 | 26.0 | 0.99 | −2.80 |
| 5572.84 | 26.0 | 3.40 | −0.28 |
| 5586.76 | 26.0 | 3.37 | −0.12 |
| 5615.64 | 26.0 | 3.33 | 0.05 |
| 6230.72 | 26.0 | 2.56 | −1.28 |
| 4178.86 | 26.1 | 2.58 | −2.51 |
| 4233.17 | 26.1 | 2.58 | −1.97 |
| 4508.29 | 26.1 | 2.86 | −2.44 |
| 4923.93 | 26.1 | 2.89 | −1.26 |
| 5018.44 | 26.1 | 2.89 | −1.10 |
| 5197.58 | 26.1 | 3.23 | −2.22 |
| 5234.63 | 26.1 | 3.22 | −2.28 |
| 6707.82 | 3.0 | 0.00 | 0.17 |
| 5889.95 | 11.0 | 0.00 | 0.12 |
| 5895.92 | 11.0 | 0.00 | −0.18 |
| 3329.92 | 12.0 | 2.71 | −1.93 |
| 3336.67 | 12.0 | 2.72 | −1.23 |
| 3986.75 | 12.0 | 4.35 | −1.44 |
| 4167.27 | 12.0 | 4.35 | −1.00 |
| 4351.91 | 12.0 | 4.35 | −0.83 |
| 5167.32 | 12.0 | 2.71 | −1.03 |
| 5172.68 | 12.0 | 2.71 | −0.40 |
| 5183.60 | 12.0 | 2.72 | −0.18 |
| 5528.41 | 12.0 | 4.35 | −0.62 |
| 3944.01 | 13.0 | 0.00 | −0.62 |
| 3961.52 | 13.0 | 0.01 | −0.32 |
| 3905.52 | 14.0 | 1.91 | −0.74 |
| 4226.73 | 20.0 | 0.00 | 0.24 |
| 4283.01 | 20.0 | 1.89 | −0.29 |
| 4289.37 | 20.0 | 1.88 | −0.39 |
| 4298.99 | 20.0 | 1.89 | −0.51 |
| 4302.53 | 20.0 | 1.90 | 0.29 |
| 4318.65 | 20.0 | 1.90 | −0.30 |
| 4425.44 | 20.0 | 1.88 | −0.36 |

**Table B.1.** Continued

| Wavelength (Å) | Species | EP (eV) | log(*gf*) (dex) |
|---|---|---|---|
| 4435.68 | 20.0 | 1.89 | −0.52 |
| 4454.78 | 20.0 | 1.90 | 0.26 |
| 4455.89 | 20.0 | 1.90 | −0.41 |
| 5265.56 | 20.0 | 2.52 | −0.15 |
| 5588.75 | 20.0 | 2.53 | 0.36 |
| 5594.46 | 20.0 | 2.52 | 0.10 |
| 5857.45 | 20.0 | 2.93 | 0.24 |
| 6102.72 | 20.0 | 1.88 | −0.79 |
| 6122.22 | 20.0 | 1.89 | −0.39 |
| 6162.17 | 20.0 | 1.90 | −0.17 |
| 6439.08 | 20.0 | 2.53 | 0.39 |
| 6493.78 | 20.0 | 2.52 | −0.11 |
| 3736.90 | 20.1 | 3.15 | −0.17 |
| 3736.90 | 20.1 | 3.15 | −0.17 |
| 3353.72 | 21.1 | 0.37 | −0.35 |
| −3353.72 | 21.1 | 0.37 | −0.52 |
| −3353.72 | 21.1 | 0.37 | −1.09 |
| −3353.73 | 21.1 | 0.37 | −0.73 |
| −3353.73 | 21.1 | 0.37 | −0.91 |
| −3353.73 | 21.1 | 0.37 | −2.09 |
| −3353.73 | 21.1 | 0.37 | −1.00 |
| −3353.73 | 21.1 | 0.37 | −0.88 |
| −3353.73 | 21.1 | 0.37 | −1.67 |
| −3353.73 | 21.1 | 0.37 | −1.42 |
| −3353.73 | 21.1 | 0.37 | −0.93 |
| −3353.73 | 21.1 | 0.37 | −1.44 |
| −3353.73 | 21.1 | 0.37 | −1.12 |
| −3353.73 | 21.1 | 0.37 | −1.29 |
| −3353.73 | 21.1 | 0.37 | −1.20 |
| 3368.93 | 21.1 | 0.01 | −2.00 |
| −3368.93 | 21.1 | 0.01 | −1.30 |
| −3368.93 | 21.1 | 0.01 | −0.81 |
| −3368.94 | 21.1 | 0.01 | −1.56 |
| −3368.94 | 21.1 | 0.01 | −1.19 |
| −3368.94 | 21.1 | 0.01 | −1.11 |
| −3368.94 | 21.1 | 0.01 | −1.29 |
| −3368.94 | 21.1 | 0.01 | −1.31 |
| −3368.94 | 21.1 | 0.01 | −1.56 |
| 3572.52 | 21.1 | 0.02 | −1.08 |
| −3572.52 | 21.1 | 0.02 | −0.28 |
| −3572.52 | 21.1 | 0.02 | −0.89 |
| −3572.52 | 21.1 | 0.02 | −0.50 |
| −3572.52 | 21.1 | 0.02 | −1.08 |
| −3572.53 | 21.1 | 0.02 | −0.84 |
| −3572.53 | 21.1 | 0.02 | −0.79 |
| −3572.53 | 21.1 | 0.02 | −0.89 |
| −3572.53 | 21.1 | 0.02 | −0.86 |
| −3572.53 | 21.1 | 0.02 | −1.18 |
| −3572.53 | 21.1 | 0.02 | −0.84 |
| −3572.53 | 21.1 | 0.02 | −0.96 |
| −3572.53 | 21.1 | 0.02 | −1.81 |
| −3572.53 | 21.1 | 0.02 | −0.86 |
| −3572.53 | 21.1 | 0.02 | −1.18 |
| −3572.53 | 21.1 | 0.02 | −0.96 |
| −3572.53 | 21.1 | 0.02 | −1.66 |
| −3572.53 | 21.1 | 0.02 | −1.18 |
| 3576.34 | 21.1 | 0.01 | −1.07 |
| −3576.34 | 21.1 | 0.01 | −0.50 |
| −3576.34 | 21.1 | 0.01 | −0.91 |
| −3576.34 | 21.1 | 0.01 | −0.89 |

**Table B.1.** Continued

| Wavelength (Å) | Species | EP (eV) | log($gf$) (dex) |
|---|---|---|---|
| −3576.34 | 21.1 | 0.01 | −1.07 |
| −3576.34 | 21.1 | 0.01 | −0.92 |
| −3576.34 | 21.1 | 0.01 | −1.59 |
| −3576.34 | 21.1 | 0.01 | −0.91 |
| −3576.34 | 21.1 | 0.01 | −1.09 |
| −3576.34 | 21.1 | 0.01 | −3.02 |
| −3576.34 | 21.1 | 0.01 | −0.92 |
| −3576.34 | 21.1 | 0.01 | −1.27 |
| −3576.34 | 21.1 | 0.01 | −1.09 |
| 3590.47 | 21.1 | 0.02 | −2.89 |
| −3590.47 | 21.1 | 0.02 | −1.89 |
| −3590.47 | 21.1 | 0.02 | −1.15 |
| −3590.47 | 21.1 | 0.02 | −2.48 |
| −3590.47 | 21.1 | 0.02 | −1.71 |
| −3590.47 | 21.1 | 0.02 | −1.33 |
| −3590.48 | 21.1 | 0.02 | −2.24 |
| −3590.48 | 21.1 | 0.02 | −1.68 |
| −3590.48 | 21.1 | 0.02 | −1.53 |
| −3590.48 | 21.1 | 0.02 | −2.10 |
| −3590.48 | 21.1 | 0.02 | −1.74 |
| −3590.48 | 21.1 | 0.02 | −1.80 |
| −3590.48 | 21.1 | 0.02 | −2.00 |
| −3590.48 | 21.1 | 0.02 | −1.92 |
| −3590.48 | 21.1 | 0.02 | −2.22 |
| 3613.82 | 21.1 | 0.02 | −0.13 |
| −3613.83 | 21.1 | 0.02 | −0.25 |
| −3613.83 | 21.1 | 0.02 | −1.06 |
| −3613.83 | 21.1 | 0.02 | −0.39 |
| −3613.83 | 21.1 | 0.02 | −0.86 |
| −3613.83 | 21.1 | 0.02 | −2.28 |
| −3613.83 | 21.1 | 0.02 | −0.55 |
| −3613.83 | 21.1 | 0.02 | −0.79 |
| −3613.83 | 21.1 | 0.02 | −1.87 |
| −3613.83 | 21.1 | 0.02 | −0.73 |
| −3613.83 | 21.1 | 0.02 | −0.80 |
| −3613.83 | 21.1 | 0.02 | −1.66 |
| −3613.84 | 21.1 | 0.02 | −0.97 |
| −3613.84 | 21.1 | 0.02 | −0.86 |
| −3613.84 | 21.1 | 0.02 | −1.55 |
| −3613.84 | 21.1 | 0.02 | −1.31 |
| −3613.84 | 21.1 | 0.02 | −0.98 |
| −3613.84 | 21.1 | 0.02 | −1.53 |
| −3613.84 | 21.1 | 0.02 | −1.16 |
| −3613.84 | 21.1 | 0.02 | −1.64 |
| 3645.30 | 21.1 | 0.02 | −2.17 |
| −3645.30 | 21.1 | 0.02 | −1.37 |
| −3645.31 | 21.1 | 0.02 | −1.98 |
| −3645.31 | 21.1 | 0.02 | −1.59 |
| −3645.31 | 21.1 | 0.02 | −2.17 |
| −3645.31 | 21.1 | 0.02 | −1.93 |
| −3645.31 | 21.1 | 0.02 | −1.88 |
| −3645.31 | 21.1 | 0.02 | −1.98 |
| −3645.32 | 21.1 | 0.02 | −1.95 |
| −3645.32 | 21.1 | 0.02 | −2.27 |
| −3645.32 | 21.1 | 0.02 | −1.93 |
| −3645.32 | 21.1 | 0.02 | −2.05 |
| −3645.32 | 21.1 | 0.02 | −2.90 |
| −3645.32 | 21.1 | 0.02 | −1.95 |
| −3645.32 | 21.1 | 0.02 | −2.27 |
| −3645.32 | 21.1 | 0.02 | −2.05 |

**Table B.1.** Continued

| Wavelength (Å) | Species | EP (eV) | log($gf$) (dex) |
|---|---|---|---|
| −3645.32 | 21.1 | 0.02 | −2.75 |
| −3645.32 | 21.1 | 0.02 | −2.27 |
| 3651.78 | 21.1 | 0.08 | −1.80 |
| −3651.78 | 21.1 | 0.08 | −1.23 |
| −3651.79 | 21.1 | 0.08 | −1.64 |
| −3651.79 | 21.1 | 0.08 | −1.62 |
| −3651.79 | 21.1 | 0.08 | −1.80 |
| −3651.80 | 21.1 | 0.08 | −1.65 |
| −3651.80 | 21.1 | 0.08 | −2.32 |
| −3651.80 | 21.1 | 0.08 | −1.64 |
| −3651.81 | 21.1 | 0.08 | −1.82 |
| −3651.81 | 21.1 | 0.08 | −3.75 |
| −3651.81 | 21.1 | 0.08 | −1.65 |
| −3651.81 | 21.1 | 0.08 | −2.00 |
| −3651.81 | 21.1 | 0.08 | −1.82 |
| 4246.81 | 21.1 | 0.32 | −0.88 |
| −4246.81 | 21.1 | 0.32 | −0.31 |
| −4246.82 | 21.1 | 0.32 | −0.72 |
| −4246.82 | 21.1 | 0.32 | −0.70 |
| −4246.82 | 21.1 | 0.32 | −0.88 |
| −4246.83 | 21.1 | 0.32 | −0.73 |
| −4246.83 | 21.1 | 0.32 | −1.40 |
| −4246.83 | 21.1 | 0.32 | −0.72 |
| −4246.83 | 21.1 | 0.32 | −0.90 |
| −4246.83 | 21.1 | 0.32 | −2.83 |
| −4246.83 | 21.1 | 0.32 | −0.73 |
| −4246.83 | 21.1 | 0.32 | −1.08 |
| −4246.83 | 21.1 | 0.32 | −0.90 |
| 4314.08 | 21.1 | 0.62 | −2.89 |
| −4314.08 | 21.1 | 0.62 | −1.68 |
| −4314.08 | 21.1 | 0.62 | −2.49 |
| −4314.08 | 21.1 | 0.62 | −0.75 |
| −4314.08 | 21.1 | 0.62 | −1.47 |
| −4314.08 | 21.1 | 0.62 | −2.28 |
| −4314.08 | 21.1 | 0.62 | −0.87 |
| −4314.08 | 21.1 | 0.62 | −2.17 |
| −4314.08 | 21.1 | 0.62 | −1.41 |
| −4314.08 | 21.1 | 0.62 | −1.41 |
| −4314.08 | 21.1 | 0.62 | −1.01 |
| −4314.09 | 21.1 | 0.62 | −2.15 |
| −4314.09 | 21.1 | 0.62 | −1.16 |
| −4314.09 | 21.1 | 0.62 | −2.25 |
| −4314.09 | 21.1 | 0.62 | −1.59 |
| −4314.09 | 21.1 | 0.62 | −1.47 |
| −4314.09 | 21.1 | 0.62 | −1.35 |
| −4314.09 | 21.1 | 0.62 | −1.78 |
| −4314.09 | 21.1 | 0.62 | −1.92 |
| −4314.09 | 21.1 | 0.62 | −1.58 |
| 4320.73 | 21.1 | 0.61 | −2.59 |
| −4320.73 | 21.1 | 0.61 | −2.18 |
| −4320.73 | 21.1 | 0.61 | −1.80 |
| −4320.73 | 21.1 | 0.61 | −1.94 |
| −4320.73 | 21.1 | 0.61 | −1.70 |
| −4320.73 | 21.1 | 0.61 | −1.59 |
| −4320.73 | 21.1 | 0.61 | −1.62 |
| −4320.73 | 21.1 | 0.61 | −1.44 |
| −4320.73 | 21.1 | 0.61 | −1.38 |
| −4320.73 | 21.1 | 0.61 | −1.41 |
| −4320.73 | 21.1 | 0.61 | −1.92 |
| −4320.73 | 21.1 | 0.61 | −1.50 |

**Table B.1.** Continued

| Wavelength (Å) | Species | EP (eV) | log($gf$) (dex) |
|---|---|---|---|
| −4320.73 | 21.1 | 0.61 | −1.23 |
| −4320.73 | 21.1 | 0.61 | −1.03 |
| −4320.73 | 21.1 | 0.61 | −0.85 |
| 4324.98 | 21.1 | 0.60 | −2.15 |
| −4324.99 | 21.1 | 0.60 | −1.71 |
| −4324.99 | 21.1 | 0.60 | −1.44 |
| −4324.99 | 21.1 | 0.60 | −1.34 |
| −4324.99 | 21.1 | 0.60 | −1.45 |
| −4324.99 | 21.1 | 0.60 | −1.46 |
| −4325.00 | 21.1 | 0.60 | −1.71 |
| −4325.00 | 21.1 | 0.60 | −1.26 |
| −4325.00 | 21.1 | 0.60 | −0.97 |
| 4374.45 | 21.1 | 0.62 | −2.10 |
| −4374.45 | 21.1 | 0.62 | −1.11 |
| −4374.45 | 21.1 | 0.62 | −1.89 |
| −4374.45 | 21.1 | 0.62 | −1.27 |
| −4374.46 | 21.1 | 0.62 | −1.82 |
| −4374.46 | 21.1 | 0.62 | −2.10 |
| −4374.46 | 21.1 | 0.62 | −1.45 |
| −4374.46 | 21.1 | 0.62 | −1.82 |
| −4374.46 | 21.1 | 0.62 | −1.89 |
| −4374.46 | 21.1 | 0.62 | −1.65 |
| −4374.46 | 21.1 | 0.62 | −1.87 |
| −4374.46 | 21.1 | 0.62 | −1.82 |
| −4374.46 | 21.1 | 0.62 | −1.87 |
| −4374.46 | 21.1 | 0.62 | −1.98 |
| −4374.46 | 21.1 | 0.62 | −2.11 |
| −4374.46 | 21.1 | 0.62 | −1.82 |
| −4374.46 | 21.1 | 0.62 | −2.21 |
| −4374.46 | 21.1 | 0.62 | −2.34 |
| −4374.46 | 21.1 | 0.62 | −1.87 |
| −4374.46 | 21.1 | 0.62 | −2.36 |
| −4374.46 | 21.1 | 0.62 | −2.21 |
| −4374.46 | 21.1 | 0.62 | −1.98 |
| 4400.38 | 21.1 | 0.61 | −2.01 |
| −4400.38 | 21.1 | 0.61 | −1.81 |
| −4400.38 | 21.1 | 0.61 | −1.20 |
| −4400.39 | 21.1 | 0.61 | −1.76 |
| −4400.39 | 21.1 | 0.61 | −1.43 |
| −4400.39 | 21.1 | 0.61 | −1.79 |
| −4400.39 | 21.1 | 0.61 | −1.72 |
| −4400.39 | 21.1 | 0.61 | −2.01 |
| −4400.39 | 21.1 | 0.61 | −1.89 |
| −4400.39 | 21.1 | 0.61 | −2.10 |
| −4400.39 | 21.1 | 0.61 | −1.81 |
| −4400.40 | 21.1 | 0.61 | −2.73 |
| −4400.40 | 21.1 | 0.61 | −2.11 |
| −4400.40 | 21.1 | 0.61 | −1.76 |
| −4400.40 | 21.1 | 0.61 | −2.59 |
| −4400.40 | 21.1 | 0.61 | −1.89 |
| −4400.40 | 21.1 | 0.61 | −1.79 |
| −4400.40 | 21.1 | 0.61 | −2.11 |
| 4415.54 | 21.1 | 0.60 | −1.86 |
| −4415.55 | 21.1 | 0.60 | −1.71 |
| −4415.55 | 21.1 | 0.60 | −1.72 |
| −4415.55 | 21.1 | 0.60 | −1.29 |
| −4415.56 | 21.1 | 0.60 | −1.89 |
| −4415.56 | 21.1 | 0.60 | −1.69 |
| −4415.56 | 21.1 | 0.60 | −2.39 |
| −4415.56 | 21.1 | 0.60 | −3.81 |

**Table B.1.** Continued

| Wavelength (Å) | Species | EP (eV) | log($gf$) (dex) |
|---|---|---|---|
| −4415.56 | 21.1 | 0.60 | −2.07 |
| −4415.57 | 21.1 | 0.60 | −1.89 |
| −4415.57 | 21.1 | 0.60 | −1.72 |
| −4415.57 | 21.1 | 0.60 | −1.71 |
| −4415.57 | 21.1 | 0.60 | −1.86 |
| 3635.46 | 22.0 | 0.00 | 0.05 |
| 3653.49 | 22.0 | 0.05 | 0.22 |
| 3729.81 | 22.0 | 0.00 | −0.35 |
| 3741.06 | 22.0 | 0.02 | −0.21 |
| 3904.78 | 22.0 | 0.90 | 0.28 |
| 3958.21 | 22.0 | 0.05 | −0.18 |
| 3989.76 | 22.0 | 0.02 | −0.20 |
| 3998.64 | 22.0 | 0.05 | −0.06 |
| 4305.91 | 22.0 | 0.85 | 0.51 |
| 4981.73 | 22.0 | 0.85 | 0.50 |
| 4991.07 | 22.0 | 0.84 | 0.38 |
| 4999.50 | 22.0 | 0.83 | 0.25 |
| 3302.10 | 22.1 | 0.15 | −2.36 |
| 3321.70 | 22.1 | 1.23 | −0.31 |
| 3335.19 | 22.1 | 0.12 | −0.42 |
| 3340.34 | 22.1 | 0.11 | −0.54 |
| 3348.84 | 22.1 | 0.12 | −1.15 |
| 3349.40 | 22.1 | 0.05 | 0.53 |
| 3372.79 | 22.1 | 0.01 | 0.28 |
| 3388.75 | 22.1 | 1.24 | −1.10 |
| 3409.81 | 22.1 | 0.03 | −1.98 |
| 3456.38 | 22.1 | 2.06 | −0.10 |
| 3491.05 | 22.1 | 0.11 | −1.15 |
| 3573.73 | 22.1 | 0.57 | −1.49 |
| 3596.05 | 22.1 | 0.61 | −1.03 |
| 3641.33 | 22.1 | 1.24 | −0.71 |
| 3659.76 | 22.1 | 1.58 | −0.53 |
| 3685.20 | 22.1 | 0.61 | 0.13 |
| 3759.29 | 22.1 | 0.61 | 0.28 |
| 3776.05 | 22.1 | 1.58 | −1.25 |
| 3813.39 | 22.1 | 0.61 | −1.83 |
| 3900.54 | 22.1 | 1.13 | −0.29 |
| 3981.99 | 22.1 | 0.57 | −2.91 |
| 4025.13 | 22.1 | 0.61 | −2.14 |
| 4028.34 | 22.1 | 1.89 | −0.92 |
| 4053.82 | 22.1 | 1.89 | −1.13 |
| 4300.04 | 22.1 | 1.18 | −0.46 |
| 4301.92 | 22.1 | 1.16 | −1.21 |
| 4312.86 | 22.1 | 1.18 | −1.12 |
| 4320.95 | 22.1 | 1.17 | −1.80 |
| 4394.06 | 22.1 | 1.22 | −1.78 |
| 4395.03 | 22.1 | 1.08 | −0.54 |
| 4399.77 | 22.1 | 1.24 | −1.19 |
| 4443.80 | 22.1 | 1.08 | −0.71 |
| 4450.48 | 22.1 | 1.08 | −1.52 |
| 4468.51 | 22.1 | 1.13 | −0.60 |
| 4501.27 | 22.1 | 1.12 | −0.77 |
| 4805.09 | 22.1 | 2.06 | −0.96 |
| 5129.16 | 22.1 | 1.89 | −1.24 |
| 5226.54 | 22.1 | 1.57 | −1.26 |
| 5336.79 | 22.1 | 1.58 | −1.59 |
| 3592.01 | 23.1 | 1.10 | −2.60 |
| −3592.02 | 23.1 | 1.10 | −2.19 |
| −3592.02 | 23.1 | 1.10 | −1.95 |
| −3592.02 | 23.1 | 1.10 | −1.61 |

**Table B.1.** Continued

| Wavelength (Å) | Species | EP (eV) | log($gf$) (dex) |
|---|---|---|---|
| −3592.02 | 23.1 | 1.10 | −1.42 |
| −3592.02 | 23.1 | 1.10 | −1.81 |
| −3592.02 | 23.1 | 1.10 | −1.39 |
| −3592.02 | 23.1 | 1.10 | −1.71 |
| −3592.02 | 23.1 | 1.10 | −1.45 |
| −3592.02 | 23.1 | 1.10 | −1.63 |
| −3592.02 | 23.1 | 1.10 | −0.87 |
| −3592.02 | 23.1 | 1.10 | −1.04 |
| −3592.02 | 23.1 | 1.10 | −1.93 |
| −3592.02 | 23.1 | 1.10 | −1.24 |
| −3592.02 | 23.1 | 1.10 | −1.51 |
| 3951.95 | 23.1 | 1.48 | −1.39 |
| −3951.95 | 23.1 | 1.48 | −2.13 |
| −3951.95 | 23.1 | 1.48 | −1.56 |
| −3951.96 | 23.1 | 1.48 | −3.13 |
| −3951.96 | 23.1 | 1.48 | −1.94 |
| −3951.96 | 23.1 | 1.48 | −1.76 |
| −3951.96 | 23.1 | 1.48 | −2.71 |
| −3951.96 | 23.1 | 1.48 | −1.91 |
| −3951.96 | 23.1 | 1.48 | −2.03 |
| −3951.97 | 23.1 | 1.48 | −2.47 |
| −3951.97 | 23.1 | 1.48 | −1.97 |
| −3951.97 | 23.1 | 1.48 | −2.45 |
| −3951.97 | 23.1 | 1.48 | −2.33 |
| −3951.97 | 23.1 | 1.48 | −2.15 |
| −3951.97 | 23.1 | 1.48 | −2.23 |
| 3578.69 | 24.0 | 0.00 | 0.41 |
| 4254.34 | 24.0 | 0.00 | −0.11 |
| 4274.80 | 24.0 | 0.00 | −0.23 |
| 4289.72 | 24.0 | 0.00 | −0.36 |
| 5206.04 | 24.0 | 0.94 | 0.02 |
| 3342.58 | 24.1 | 2.46 | −0.74 |
| 3358.49 | 24.1 | 2.46 | −0.59 |
| 3382.68 | 24.1 | 2.46 | −0.95 |
| 3408.76 | 24.1 | 2.48 | −0.39 |
| 3315.66 | 28.0 | 0.11 | −1.23 |
| 4030.73 | 25.0 | 0.00 | −1.04 |
| −4030.75 | 25.0 | 0.00 | −1.96 |
| −4030.75 | 25.0 | 0.00 | −1.18 |
| −4030.76 | 25.0 | 0.00 | −3.17 |
| −4030.76 | 25.0 | 0.00 | −1.78 |
| −4030.76 | 25.0 | 0.00 | −1.34 |
| −4030.77 | 25.0 | 0.00 | −2.82 |
| −4030.77 | 25.0 | 0.00 | −1.75 |
| −4030.77 | 25.0 | 0.00 | −1.52 |
| −4030.78 | 25.0 | 0.00 | −2.70 |
| −4030.78 | 25.0 | 0.00 | −1.82 |
| −4030.78 | 25.0 | 0.00 | −1.74 |
| −4030.78 | 25.0 | 0.00 | −2.00 |
| −4030.78 | 25.0 | 0.00 | −2.77 |
| −4030.78 | 25.0 | 0.00 | −2.03 |
| 4033.04 | 25.0 | 0.00 | −1.20 |
| −4033.05 | 25.0 | 0.00 | −1.98 |
| −4033.06 | 25.0 | 0.00 | −1.98 |
| −4033.06 | 25.0 | 0.00 | −1.46 |
| −4033.06 | 25.0 | 0.00 | −1.82 |
| −4033.07 | 25.0 | 0.00 | −1.82 |
| −4033.07 | 25.0 | 0.00 | −1.79 |
| −4033.07 | 25.0 | 0.00 | −1.81 |
| −4033.08 | 25.0 | 0.00 | −1.81 |

**Table B.1.** Continued

| Wavelength (Å) | Species | EP (eV) | log($gf$) (dex) |
|---|---|---|---|
| −4033.08 | 25.0 | 0.00 | −2.24 |
| −4033.08 | 25.0 | 0.00 | −1.91 |
| −4033.08 | 25.0 | 0.00 | −1.91 |
| −4033.09 | 25.0 | 0.00 | −2.94 |
| −4033.09 | 25.0 | 0.00 | −2.17 |
| −4033.09 | 25.0 | 0.00 | −2.17 |
| 4034.47 | 25.0 | 0.00 | −1.33 |
| −4034.47 | 25.0 | 0.00 | −2.02 |
| −4034.47 | 25.0 | 0.00 | −2.97 |
| −4034.48 | 25.0 | 0.00 | −1.54 |
| −4034.49 | 25.0 | 0.00 | −1.87 |
| −4034.49 | 25.0 | 0.00 | −2.59 |
| −4034.49 | 25.0 | 0.00 | −1.81 |
| −4034.50 | 25.0 | 0.00 | −2.41 |
| −4034.50 | 25.0 | 0.00 | −1.89 |
| −4034.50 | 25.0 | 0.00 | −2.22 |
| −4034.50 | 25.0 | 0.00 | −2.37 |
| −4034.50 | 25.0 | 0.00 | −2.05 |
| 3405.07 | 27.0 | 0.43 | −1.59 |
| −3405.08 | 27.0 | 0.43 | −1.39 |
| −3405.08 | 27.0 | 0.43 | −1.37 |
| −3405.08 | 27.0 | 0.43 | −1.38 |
| −3405.08 | 27.0 | 0.43 | −1.26 |
| −3405.08 | 27.0 | 0.43 | −1.59 |
| −3405.09 | 27.0 | 0.43 | −1.24 |
| −3405.09 | 27.0 | 0.43 | −1.22 |
| −3405.09 | 27.0 | 0.43 | −1.37 |
| −3405.09 | 27.0 | 0.43 | −1.07 |
| −3405.10 | 27.0 | 0.43 | −1.23 |
| −3405.10 | 27.0 | 0.43 | −1.26 |
| −3405.10 | 27.0 | 0.43 | −0.90 |
| −3405.11 | 27.0 | 0.43 | −1.31 |
| −3405.11 | 27.0 | 0.43 | −1.22 |
| −3405.12 | 27.0 | 0.43 | −0.74 |
| −3405.12 | 27.0 | 0.43 | −1.52 |
| −3405.12 | 27.0 | 0.43 | −1.23 |
| −3405.13 | 27.0 | 0.43 | −0.60 |
| −3405.14 | 27.0 | 0.43 | −1.31 |
| −3405.15 | 27.0 | 0.43 | −0.46 |
| −3405.16 | 27.0 | 0.43 | −1.52 |
| 3412.32 | 27.0 | 0.51 | −1.51 |
| −3412.32 | 27.0 | 0.51 | −1.31 |
| −3412.33 | 27.0 | 0.51 | −1.78 |
| −3412.33 | 27.0 | 0.51 | −1.14 |
| −3412.33 | 27.0 | 0.51 | −1.71 |
| −3412.33 | 27.0 | 0.51 | −0.99 |
| −3412.33 | 27.0 | 0.51 | −1.51 |
| −3412.33 | 27.0 | 0.51 | −0.86 |
| −3412.33 | 27.0 | 0.51 | −2.41 |
| −3412.33 | 27.0 | 0.51 | −1.41 |
| −3412.33 | 27.0 | 0.51 | −0.75 |
| −3412.34 | 27.0 | 0.51 | −1.36 |
| −3412.34 | 27.0 | 0.51 | −2.25 |
| −3412.34 | 27.0 | 0.51 | −0.64 |
| −3412.34 | 27.0 | 0.51 | −1.37 |
| −3412.34 | 27.0 | 0.51 | −2.25 |
| −3412.35 | 27.0 | 0.51 | −1.44 |
| −3412.35 | 27.0 | 0.51 | −2.35 |
| −3412.35 | 27.0 | 0.51 | −1.65 |
| −3412.36 | 27.0 | 0.51 | −2.55 |

**Table B.1.** Continued

| Wavelength (Å) | Species | EP (eV) | log($gf$) (dex) |
|---|---|---|---|
| −3412.36 | 27.0 | 0.51 | −2.95 |
| 3412.59 | 27.0 | 0.00 | −3.76 |
| −3412.60 | 27.0 | 0.00 | −3.36 |
| −3412.61 | 27.0 | 0.00 | −2.46 |
| −3412.61 | 27.0 | 0.00 | −3.16 |
| −3412.62 | 27.0 | 0.00 | −2.25 |
| −3412.62 | 27.0 | 0.00 | −3.06 |
| −3412.62 | 27.0 | 0.00 | −1.45 |
| −3412.62 | 27.0 | 0.00 | −2.18 |
| −3412.63 | 27.0 | 0.00 | −3.06 |
| −3412.63 | 27.0 | 0.00 | −1.56 |
| −3412.63 | 27.0 | 0.00 | −2.17 |
| −3412.63 | 27.0 | 0.00 | −2.22 |
| −3412.63 | 27.0 | 0.00 | −1.67 |
| −3412.63 | 27.0 | 0.00 | −3.22 |
| −3412.64 | 27.0 | 0.00 | −2.32 |
| −3412.64 | 27.0 | 0.00 | −1.80 |
| −3412.64 | 27.0 | 0.00 | −2.52 |
| −3412.64 | 27.0 | 0.00 | −1.95 |
| −3412.64 | 27.0 | 0.00 | −2.59 |
| −3412.64 | 27.0 | 0.00 | −2.12 |
| −3412.64 | 27.0 | 0.00 | −2.32 |
| 3431.55 | 27.0 | 0.10 | −3.52 |
| −3431.56 | 27.0 | 0.10 | −3.11 |
| −3431.56 | 27.0 | 0.10 | −2.41 |
| −3431.57 | 27.0 | 0.10 | −2.89 |
| −3431.57 | 27.0 | 0.10 | −2.21 |
| −3431.57 | 27.0 | 0.10 | −2.77 |
| −3431.57 | 27.0 | 0.10 | −2.16 |
| −3431.58 | 27.0 | 0.10 | −1.57 |
| −3431.58 | 27.0 | 0.10 | −2.72 |
| −3431.58 | 27.0 | 0.10 | −2.18 |
| −3431.58 | 27.0 | 0.10 | −1.71 |
| −3431.58 | 27.0 | 0.10 | −2.75 |
| −3431.58 | 27.0 | 0.10 | −2.27 |
| −3431.58 | 27.0 | 0.10 | −2.46 |
| −3431.58 | 27.0 | 0.10 | −1.88 |
| −3431.59 | 27.0 | 0.10 | −2.07 |
| −3431.59 | 27.0 | 0.10 | −2.72 |
| −3431.59 | 27.0 | 0.10 | −2.33 |
| 3433.04 | 27.0 | 0.63 | −1.33 |
| −3433.04 | 27.0 | 0.63 | −1.16 |
| −3433.04 | 27.0 | 0.63 | −1.16 |
| −3433.04 | 27.0 | 0.63 | −1.03 |
| −3433.04 | 27.0 | 0.63 | −1.03 |
| −3433.04 | 27.0 | 0.63 | −1.35 |
| −3433.04 | 27.0 | 0.63 | −1.15 |
| −3433.04 | 27.0 | 0.63 | −1.15 |
| −3433.04 | 27.0 | 0.63 | −0.74 |
| 3449.14 | 27.0 | 0.58 | −1.74 |
| −3449.14 | 27.0 | 0.58 | −1.49 |
| −3449.15 | 27.0 | 0.58 | −1.49 |
| −3449.15 | 27.0 | 0.58 | −3.39 |
| −3449.15 | 27.0 | 0.58 | −1.29 |
| −3449.15 | 27.0 | 0.58 | −1.29 |
| −3449.15 | 27.0 | 0.58 | −2.15 |
| −3449.15 | 27.0 | 0.58 | −1.22 |
| −3449.16 | 27.0 | 0.58 | −1.22 |
| −3449.16 | 27.0 | 0.58 | −1.43 |
| −3449.16 | 27.0 | 0.58 | −1.25 |

**Table B.1.** Continued

| Wavelength (Å) | Species | EP (eV) | log($gf$) (dex) |
|---|---|---|---|
| −3449.17 | 27.0 | 0.58 | −1.25 |
| −3449.17 | 27.0 | 0.58 | −1.03 |
| −3449.17 | 27.0 | 0.58 | −1.44 |
| −3449.19 | 27.0 | 0.58 | −1.44 |
| −3449.19 | 27.0 | 0.58 | −0.74 |
| 3449.38 | 27.0 | 0.43 | −2.14 |
| −3449.38 | 27.0 | 0.43 | −2.34 |
| −3449.38 | 27.0 | 0.43 | −2.34 |
| −3449.38 | 27.0 | 0.43 | −2.13 |
| −3449.38 | 27.0 | 0.43 | −2.12 |
| −3449.39 | 27.0 | 0.43 | −2.12 |
| −3449.39 | 27.0 | 0.43 | −1.99 |
| −3449.39 | 27.0 | 0.43 | −2.01 |
| −3449.41 | 27.0 | 0.43 | −2.01 |
| −3449.41 | 27.0 | 0.43 | −1.82 |
| −3449.41 | 27.0 | 0.43 | −1.97 |
| −3449.42 | 27.0 | 0.43 | −1.97 |
| −3449.42 | 27.0 | 0.43 | −1.65 |
| −3449.42 | 27.0 | 0.43 | −1.98 |
| −3449.44 | 27.0 | 0.43 | −1.98 |
| −3449.44 | 27.0 | 0.43 | −1.49 |
| −3449.44 | 27.0 | 0.43 | −2.06 |
| −3449.47 | 27.0 | 0.43 | −2.06 |
| −3449.47 | 27.0 | 0.43 | −1.35 |
| −3449.47 | 27.0 | 0.43 | −2.27 |
| −3449.49 | 27.0 | 0.43 | −2.27 |
| −3449.49 | 27.0 | 0.43 | −1.21 |
| 3453.47 | 27.0 | 0.43 | −1.05 |
| −3453.48 | 27.0 | 0.43 | −0.93 |
| −3453.48 | 27.0 | 0.43 | −1.50 |
| −3453.48 | 27.0 | 0.43 | −0.81 |
| −3453.49 | 27.0 | 0.43 | −1.29 |
| −3453.49 | 27.0 | 0.43 | −0.70 |
| −3453.49 | 27.0 | 0.43 | −2.50 |
| −3453.49 | 27.0 | 0.43 | −1.19 |
| −3453.50 | 27.0 | 0.43 | −0.59 |
| −3453.50 | 27.0 | 0.43 | −2.29 |
| −3453.51 | 27.0 | 0.43 | −1.16 |
| −3453.51 | 27.0 | 0.43 | −0.50 |
| −3453.51 | 27.0 | 0.43 | −2.26 |
| −3453.52 | 27.0 | 0.43 | −1.17 |
| −3453.52 | 27.0 | 0.43 | −0.41 |
| −3453.52 | 27.0 | 0.43 | −2.34 |
| −3453.53 | 27.0 | 0.43 | −1.25 |
| −3453.54 | 27.0 | 0.43 | −0.32 |
| −3453.54 | 27.0 | 0.43 | −2.54 |
| −3453.55 | 27.0 | 0.43 | −1.47 |
| −3453.56 | 27.0 | 0.43 | −2.92 |
| 3594.83 | 27.0 | 0.17 | −2.62 |
| −3594.83 | 27.0 | 0.17 | −2.37 |
| −3594.84 | 27.0 | 0.17 | −2.37 |
| −3594.84 | 27.0 | 0.17 | −4.27 |
| −3594.84 | 27.0 | 0.17 | −2.17 |
| −3594.85 | 27.0 | 0.17 | −2.17 |
| −3594.85 | 27.0 | 0.17 | −3.03 |
| −3594.85 | 27.0 | 0.17 | −2.10 |
| −3594.86 | 27.0 | 0.17 | −2.10 |
| −3594.86 | 27.0 | 0.17 | −2.31 |
| −3594.86 | 27.0 | 0.17 | −2.13 |
| −3594.87 | 27.0 | 0.17 | −2.13 |

**Table B.1.** Continued

| Wavelength (Å) | Species | EP (eV) | log($gf$) (dex) |
|---|---|---|---|
| −3594.87 | 27.0 | 0.17 | −1.91 |
| −3594.87 | 27.0 | 0.17 | −2.32 |
| −3594.89 | 27.0 | 0.17 | −2.32 |
| −3594.89 | 27.0 | 0.17 | −1.62 |
| 3845.45 | 27.0 | 0.92 | −0.66 |
| −3845.46 | 27.0 | 0.92 | −0.77 |
| −3845.46 | 27.0 | 0.92 | −0.88 |
| −3845.46 | 27.0 | 0.92 | −1.01 |
| −3845.47 | 27.0 | 0.92 | −1.16 |
| −3845.47 | 27.0 | 0.92 | −1.33 |
| −3845.47 | 27.0 | 0.92 | −1.67 |
| −3845.47 | 27.0 | 0.92 | −1.53 |
| −3845.47 | 27.0 | 0.92 | −1.46 |
| −3845.47 | 27.0 | 0.92 | −1.39 |
| −3845.47 | 27.0 | 0.92 | −1.80 |
| −3845.47 | 27.0 | 0.92 | −1.38 |
| −3845.47 | 27.0 | 0.92 | −1.43 |
| −3845.47 | 27.0 | 0.92 | −1.73 |
| −3845.47 | 27.0 | 0.92 | −1.53 |
| −3845.48 | 27.0 | 0.92 | −2.43 |
| −3845.48 | 27.0 | 0.92 | −2.27 |
| −3845.48 | 27.0 | 0.92 | −2.27 |
| −3845.48 | 27.0 | 0.92 | −2.37 |
| −3845.48 | 27.0 | 0.92 | −2.57 |
| −3845.48 | 27.0 | 0.92 | −2.97 |
| 3873.07 | 27.0 | 0.43 | −3.64 |
| −3873.07 | 27.0 | 0.43 | −3.24 |
| −3873.08 | 27.0 | 0.43 | −3.04 |
| −3873.08 | 27.0 | 0.43 | −2.94 |
| −3873.08 | 27.0 | 0.43 | −2.94 |
| −3873.08 | 27.0 | 0.43 | −3.10 |
| −3873.09 | 27.0 | 0.43 | −2.40 |
| −3873.09 | 27.0 | 0.43 | −2.20 |
| −3873.09 | 27.0 | 0.43 | −2.10 |
| −3873.09 | 27.0 | 0.43 | −2.47 |
| −3873.09 | 27.0 | 0.43 | −2.05 |
| −3873.10 | 27.0 | 0.43 | −2.06 |
| −3873.10 | 27.0 | 0.43 | −2.20 |
| −3873.10 | 27.0 | 0.43 | −2.13 |
| −3873.10 | 27.0 | 0.43 | −2.34 |
| −3873.10 | 27.0 | 0.43 | −2.00 |
| −3873.11 | 27.0 | 0.43 | −1.83 |
| −3873.11 | 27.0 | 0.43 | −1.68 |
| −3873.12 | 27.0 | 0.43 | −1.55 |
| −3873.13 | 27.0 | 0.43 | −1.44 |
| −3873.13 | 27.0 | 0.43 | −1.33 |
| 3995.27 | 27.0 | 0.92 | −2.03 |
| −3995.27 | 27.0 | 0.92 | −1.96 |
| −3995.27 | 27.0 | 0.92 | −1.76 |
| −3995.28 | 27.0 | 0.92 | −2.66 |
| −3995.28 | 27.0 | 0.92 | −1.76 |
| −3995.28 | 27.0 | 0.92 | −1.56 |
| −3995.28 | 27.0 | 0.92 | −2.50 |
| −3995.28 | 27.0 | 0.92 | −1.66 |
| −3995.28 | 27.0 | 0.92 | −1.39 |
| −3995.29 | 27.0 | 0.92 | −2.50 |
| −3995.29 | 27.0 | 0.92 | −1.61 |
| −3995.29 | 27.0 | 0.92 | −1.24 |
| −3995.30 | 27.0 | 0.92 | −2.60 |
| −3995.30 | 27.0 | 0.92 | −1.62 |

**Table B.1.** Continued

| Wavelength (Å) | Species | EP (eV) | log($gf$) (dex) |
|---|---|---|---|
| −3995.30 | 27.0 | 0.92 | −1.11 |
| −3995.31 | 27.0 | 0.92 | −2.80 |
| −3995.31 | 27.0 | 0.92 | −1.69 |
| −3995.31 | 27.0 | 0.92 | −1.00 |
| −3995.33 | 27.0 | 0.92 | −3.20 |
| −3995.33 | 27.0 | 0.92 | −1.90 |
| −3995.33 | 27.0 | 0.92 | −0.89 |
| 4121.29 | 27.0 | 0.92 | −0.99 |
| −4121.30 | 27.0 | 0.92 | −1.10 |
| −4121.31 | 27.0 | 0.92 | −1.21 |
| −4121.31 | 27.0 | 0.92 | −1.34 |
| −4121.32 | 27.0 | 0.92 | −2.00 |
| −4121.32 | 27.0 | 0.92 | −1.49 |
| −4121.32 | 27.0 | 0.92 | −1.79 |
| −4121.32 | 27.0 | 0.92 | −1.66 |
| −4121.32 | 27.0 | 0.92 | −1.72 |
| −4121.33 | 27.0 | 0.92 | −1.86 |
| −4121.33 | 27.0 | 0.92 | −1.71 |
| −4121.33 | 27.0 | 0.92 | −2.13 |
| −4121.33 | 27.0 | 0.92 | −1.76 |
| −4121.33 | 27.0 | 0.92 | −1.86 |
| −4121.33 | 27.0 | 0.92 | −2.06 |
| −4121.34 | 27.0 | 0.92 | −2.76 |
| −4121.34 | 27.0 | 0.92 | −3.30 |
| −4121.34 | 27.0 | 0.92 | −2.60 |
| −4121.34 | 27.0 | 0.92 | −2.60 |
| −4121.34 | 27.0 | 0.92 | −2.70 |
| −4121.34 | 27.0 | 0.92 | −2.90 |
| 3365.76 | 28.0 | 0.42 | −1.19 |
| 3380.57 | 28.0 | 0.42 | −0.17 |
| 3380.87 | 28.0 | 0.28 | −1.34 |
| 3391.04 | 28.0 | 0.00 | −1.05 |
| 3452.89 | 28.0 | 0.11 | −0.91 |
| 3458.46 | 28.0 | 0.21 | −0.22 |
| 3461.65 | 28.0 | 0.03 | −0.35 |
| 3472.54 | 28.0 | 0.11 | −0.81 |
| 3492.95 | 28.0 | 0.11 | −0.25 |
| 3587.93 | 28.0 | 0.03 | −2.34 |
| 3597.70 | 28.0 | 0.21 | −1.10 |
| 3610.46 | 28.0 | 0.11 | −1.15 |
| 3612.73 | 28.0 | 0.28 | −1.41 |
| 3619.39 | 28.0 | 0.42 | 0.04 |
| 3775.57 | 28.0 | 0.42 | −1.39 |
| 3783.52 | 28.0 | 0.42 | −1.31 |
| 3807.14 | 28.0 | 0.42 | −1.21 |
| 3858.29 | 28.0 | 0.42 | −0.94 |
| 5476.90 | 28.0 | 1.83 | −0.89 |
| 4810.53 | 30.0 | 4.08 | −0.31 |
| 4077.71 | 38.1 | 0.00 | 0.17 |
| 4215.52 | 38.1 | 0.00 | −0.15 |
| 3600.74 | 39.1 | 0.18 | 0.28 |
| 3611.04 | 39.1 | 0.13 | 0.11 |
| 3710.29 | 39.1 | 0.18 | 0.46 |
| 3774.33 | 39.1 | 0.13 | 0.21 |
| 3788.69 | 39.1 | 0.10 | −0.07 |
| 3991.13 | 40.1 | 0.76 | −0.31 |
| 3998.97 | 40.1 | 0.56 | −0.52 |
| 4149.20 | 40.1 | 0.80 | −0.04 |
| 4934.10 | 56.1 | 0.00 | −1.77 |
| −4934.06 | 56.1 | 0.00 | −1.84 |

**Table B.1.** Continued

| Wavelength (Å) | Species | EP (eV) | log($gf$) (dex) |
|---|---|---|---|
| −4934.07 | 56.1 | 0.00 | −2.54 |
| −4934.12 | 56.1 | 0.00 | −1.84 |
| −4934.13 | 56.1 | 0.00 | −1.84 |
| −4934.10 | 56.1 | 0.00 | −1.26 |
| −4934.05 | 56.1 | 0.00 | −1.61 |
| −4934.07 | 56.1 | 0.00 | −2.30 |
| −4934.12 | 56.1 | 0.00 | −1.61 |
| −4934.13 | 56.1 | 0.00 | −1.61 |
| −4934.10 | 56.1 | 0.00 | −0.29 |
| 6141.70 | 56.1 | 0.70 | −3.63 |
| −6141.70 | 56.1 | 0.70 | −3.40 |
| −6141.70 | 56.1 | 0.70 | −2.49 |
| −6141.70 | 56.1 | 0.70 | −2.26 |
| −6141.70 | 56.1 | 0.70 | −3.46 |
| −6141.70 | 56.1 | 0.70 | −3.22 |
| −6141.70 | 56.1 | 0.70 | −1.68 |
| −6141.70 | 56.1 | 0.70 | −2.39 |
| −6141.70 | 56.1 | 0.70 | −2.16 |
| −6141.70 | 56.1 | 0.70 | −1.45 |
| −6141.70 | 56.1 | 0.70 | −1.70 |
| −6141.70 | 56.1 | 0.70 | −1.18 |
| −6141.70 | 56.1 | 0.70 | −0.22 |
| −6141.70 | 56.1 | 0.70 | −2.51 |
| −6141.70 | 56.1 | 0.70 | −1.89 |
| −6141.70 | 56.1 | 0.70 | −2.27 |
| −6141.70 | 56.1 | 0.70 | −1.66 |
| −6141.70 | 56.1 | 0.70 | −2.46 |
| −6141.70 | 56.1 | 0.70 | −2.14 |
| −6141.70 | 56.1 | 0.70 | −1.90 |
| −6141.70 | 56.1 | 0.70 | −2.23 |
| 6496.90 | 56.1 | 0.60 | −2.00 |
| −6496.91 | 56.1 | 0.60 | −2.76 |
| −6496.91 | 56.1 | 0.60 | −2.37 |
| −6496.89 | 56.1 | 0.60 | −3.07 |
| −6496.91 | 56.1 | 0.60 | −2.37 |
| −6496.89 | 56.1 | 0.60 | −2.37 |
| −6496.90 | 56.1 | 0.60 | −1.92 |
| −6496.90 | 56.1 | 0.60 | −1.48 |
| −6496.91 | 56.1 | 0.60 | −2.53 |
| −6496.91 | 56.1 | 0.60 | −2.13 |
| −6496.89 | 56.1 | 0.60 | −2.83 |
| −6496.91 | 56.1 | 0.60 | −2.13 |
| −6496.89 | 56.1 | 0.60 | −2.13 |
| −6496.90 | 56.1 | 0.60 | −1.69 |
| −6496.90 | 56.1 | 0.60 | −0.52 |